\RequirePackage{fix-cm}
\documentclass[11pt]{article}
\usepackage[fontsize=10.8pt]{scrextend}

\usepackage[round]{natbib}
\RequirePackage[OT1]{fontenc}
\RequirePackage{amsthm,amsmath}
\allowdisplaybreaks[4]
\usepackage{array}
\usepackage{amssymb,bbm}
\usepackage{amsfonts}
\usepackage{enumerate}
\usepackage{fullpage}	
\usepackage{amscd}
\usepackage{color}
\usepackage{mathrsfs}
\usepackage{bbm}
\usepackage{bm}
\usepackage{float}
\usepackage{threeparttable}
\usepackage{rotating}
\usepackage{graphicx}
 \usepackage[titletoc,toc,title]{appendix}
\usepackage[ruled,linesnumbered]{algorithm2e}
\usepackage{xr-hyper}
\usepackage{multirow}
\usepackage{hyperref}[]
\hypersetup{
	colorlinks=true,
	linkcolor=blue,
	filecolor=magenta,      
	urlcolor=cyan,
	citecolor=blue,
}
\usepackage{cleveref}
\usepackage{subcaption}
\usepackage[hmargin={0.8in, 0.8in}, vmargin={0.75in, 0.75in}]{geometry}
\usepackage{booktabs}
\usepackage{array}
\newcolumntype{Z}{>{\setbox0=\hbox\bgroup}c<{\egroup}@{\hspace*{-\tabcolsep}}}
\usepackage{setspace}
\usepackage{diagbox}
\usepackage{tabu}
\usepackage{titlesec}
\usepackage{authblk}
\titlespacing*\section{0pt}{6pt plus 2pt minus 2pt}{0pt plus 2pt minus 2pt}
\titlespacing*\subsection{0pt}{4pt plus 2pt minus 2pt}{0pt plus 2pt minus 2pt}
\titlespacing*\subsubsection{0pt}{2pt plus 2pt minus 2pt}{0pt plus 2pt minus 2pt}

\newcommand{\E}{\mathbb{E}}
\newcommand{\floor}[1]{\lfloor#1\rfloor}
\newcommand{\blind}{1}
\theoremstyle{it}
\newtheorem{thm}{Theorem}[section]
\newtheorem{lemma}{Lemma}[section]
\newtheorem{ass}{Assumption}[section]

\newtheorem{defn}{Definition}[section]
\SetKwInput{KwIni}{Initialization}
\SetKwInput{KwPro}{Procedure}

\makeatletter
\newcommand*{\addFileDependency}[1]{
  \typeout{(#1)}
  \@addtofilelist{#1}
  \IfFileExists{#1}{}{\typeout{No file #1.}}
}
\newcommand*\showfontsize{\f@size{} point}
\makeatother

\newcommand*{\myexternaldocument}[1]{%
    \externaldocument{#1}%
    \addFileDependency{#1.tex}%
    \addFileDependency{#1.aux}%
}
\myexternaldocument{JMA_Supplement}
\date{\vspace{-5ex}}
\begin{document}

\def\spacingset#1{\renewcommand{\baselinestretch}%
{#1}\small\normalsize} \spacingset{1}

\if1\blind
{
  \title{Robust Inference for Change Points in High Dimension }

    \author[1]{Feiyu Jiang}
    \author[2]{Runmin Wang}
    \author[3]{Xiaofeng Shao}
    
    \affil[1]{Department of Statistics and Data Science\\ School of Management, Fudan University}
    \affil[2]{Department of Statistical Science\\ Southern Methodist  University}
    \affil[3]{Department of Statistics\\
    University of Illinois at Urbana Champaign}
		
	\date{}	
	\maketitle
} \fi

\if0\blind
{
	\title{Robust Inference for Change Points in High Dimension}	
	\author{}
	\date{}
	\maketitle
} \fi

\bigskip
\baselineskip=2\baselineskip
\begin{abstract}
	\baselineskip=1.25\baselineskip
		This paper proposes a new test for a change point in the mean of high-dimensional data based on the spatial sign and self-normalization. The test is  easy to implement with no tuning parameters,  robust to heavy-tailedness and  theoretically justified with both fixed-$n$ and sequential asymptotics under both null and   alternatives, where $n$ is the sample size.  We demonstrate that the fixed-$n$ asymptotics provide a better approximation to the finite sample distribution and thus should be preferred in both testing and testing-based estimation. To estimate  the number and locations when multiple change-points are present, we propose to combine the p-value under the fixed-$n$ asymptotics  with the seeded binary segmentation (SBS) algorithm. Through numerical experiments, we show that the spatial sign based procedures are robust with respect to the heavy-tailedness and strong coordinate-wise dependence, whereas their non-robust counterparts proposed in
\cite{wang2021inference} appear to under-perform.  A real data example is also provided to  illustrate the robustness and broad applicability of the proposed test and its corresponding estimation algorithm. 
\end{abstract}

\noindent%
{\it Keywords:}  Change Points, High Dimensional Data, Segmentation, Self-Normalization, Spatial Sign

\setcounter{section}{0} 
\setcounter{equation}{0} 

\spacingset{1.25} 
\section{Introduction}

High-dimensional data analysis often encounters testing and estimation of change-points in the mean, and it has attracted a lot of attention in statistics recently. See \cite{horvath2012change}, \cite{jirak2015uniform}, \cite{cho2016change}, \cite{wang2018high}, \cite{liu2020unified}, \cite{wang2021inference}, \cite{zhang2021adaptive} and \cite{yu2021finite,yu2022robust}  for some recent literature. Among the proposed tests and estimation methods, most of them require quite strong moment conditions (e.g., Gaussian or sub-Gaussian assumption, or sixth moment assumption) and some of them also require weak component-wise dependence assumption. There are only a few exceptions, such as \cite{yu2022robust}, where they used   anti-symmetric and
nonlinear kernels in a U-statistics framework   to achieve robustness. However, the limiting distribution of their test statistic is non-pivotal and their procedure requires bootstrap calibration, which could be computationally demanding. In addition, their test statistic targets the sparse alternative only. As pointed out in athe review paper by \cite{liu2021high}, the interest in the dense alternative can be well motivated by real data and is often the type of alternative the practitioners want to detect.  For example, copy
number variations in cancer cells are commonly manifested as change-points occurring at the same 
positions across many related data sequences corresponding to cancer samples and biologically related individuals; see \cite{fan2017}.

In this article, we propose a new test for a change point in the mean of high-dimensional data that works for a broad class of data generating processes. In particular, our test targets the dense alternative, is robust to heavy-tailedness, and can accommodate both weak and strong coordinate-wise dependence. Our test is built on two recent advances in high-dimensional testing: spatial sign based two sample test developed in \cite{chakraborty2017tests} and U-statistics based change-point test developed in \cite{wang2021inference}. 
Spatial sign based tests have been studied in the literature of multivariate data   and they are usually used to handle heavy-tailedness, see \cite{oja2010multivariate} for a book-length review. However, it was until recently that \cite{wang2015high} and \cite{chakraborty2017tests} discovered that spatial sign could also help relax the restrictive moment conditions in high dimensional testing problems. In \cite{wang2021inference}, they advanced the high-dimensional two sample U-statistic pioneered by \cite{chen2010two} to the change-point setting by adopting the self-normalization (SN) \citep{shao2010self,shao2010testing}. Their test targets dense alternative, but requires sixth moment assumption and only allows for weak coordinate-wise dependence.

Building on these two recent advances, we shall propose a spatial signed SN-based test for a change point in the mean of high-dimensional data. 
Our contribution  to the literature is threefold. Firstly,  we derive the limiting null distribution of our test statistic under the so-called fixed-$n$ asymptotics, where the sample size $n$ is fixed and dimension $p$ grows to infinity. We discovered that the fixed-$n$ asymptotics provide a better approximation to the finite sample distribution when the sample size is small or moderate. We also let $n$ grows to infinity after we derive $n$-dependent asymptotic distribution, and obtain the limit under the sequential asymptotics \citep{phillips1999}. This type of asymptotics seems new to the high-dimensional change-point literature and may  be more broadly adopted in change-point testing and other high-dimensional problems. Secondly, our asymptotic theory covers both scenarios,  the weak coordinate-wise dependence via $\rho$ mixing, and strong coordindate-wise dependence under the framework of ``randomly scaled $\rho$-mixing sequence" (RSRM) in \cite{chakraborty2017tests}. The process convergence associated with spatial signed U-process we develop in this paper further facilitates the application of our  test under sequential asymptotics where $n$, in addition to $p$, also goes to infinity.  In particular, we have developed novel theory to establish the process convergence result under the RSRM framework. In general, this requires to show the finite dimensional convergence and asymptotic equicontinuity (tightness). For the tightness, we derive a bound for the eighth moment of the increment of the sample path based on a conditional   argument under the sequential asymptotics, which is new to the literature. Using this new technique, we provide the unconditional limiting null distribution of the test statistic for the fixed-$n$ and growing-$p$ case. This is    stronger than the results in \cite{chakraborty2017tests} which is a conditional limiting null distribution.  Thirdly, We extend our test to estimate multiple changes by combining the p-value  based on the fixed-$n$ asymptotics and the seeded binary segmentation (SBS) \citep{kovacs2020seeded}. The use of fixed-$n$ asymptotics is especially recommended due to the fact that in these popular  generic segmentation algorithms such as WBS \citep{fryzlewicz2014wild} and SBS, test statistics over many intervals of small/moderate lengths are calculated and the sequential asymptotics is not accurate in approximating the finite sample distribution, as compared to its fixed-$n$ counterpart. The superiority and robustness of our estimation algorithm is corroborated in a small simulation study.

The rest of the paper is organized as follows. In Section \ref{sec:test}, we define the spatial signed SN test. Section \ref{sec:theory} studies the asymptotic behavior of the test under both the null and  local alternatives.  Extensions to estimating multiple change-points  are
elaborated in Section \ref{sec:est}.  Numerical studies for both testing and estimation are relegated to Section \ref{sec:num}. Section \ref{sec:real} contains a real data example and Section \ref{sec:con} concludes.  All proofs with auxiliary lemmas are given in the appendix. 
Throughout the paper, we  denote  $\to_p$ as the convergence in probability, $\overset{\mathcal{D}}{\rightarrow}$ as  the convergence in distribution and  $\rightsquigarrow$ as the
weak convergence for stochastic processes. The notations $\bm{1}_d$ and $\bm{0}_d$  are used to represent vectors of dimension $d$ whose entries are all ones and zeros, respectively.  For $a,b\in\mathbb{R}$, denote $a\wedge b=\min(a,b)$ and $a\vee b=\max(a,b)$.  For a vector $a\in \mathbb{R}^d$, $\|a\|$ denotes its Euclidean norm.  For a matrix $A$, $\|A\|_F$ denotes its Frobenius norm.

\section{Test Statistics}\label{sec:test}
Let $\{X_i\}_{i=1}^n$ be a sequence of i.i.d $\mathbb{R}^p$-valued random vectors with mean $0$ and covariance $\Sigma$. We assume that  the observed data $\{Y_i\}_{i=1}^n$ satisfies $Y_i=\mu_i+X_i$, where $\mathbb{E}X_i=\bm{0}_p$ and $\mu_i\in\mathbb{R}^p$ is the mean at time $i$. We are interested in the following testing problem: 
\begin{equation}\label{test}
H_0: \mu_1=\cdots=\mu_n,\quad\text{v.s.}\quad H_1:\mu_1=\cdots=\mu_{k^*}\neq \mu_{k^*+1}=\cdots=\mu_n,\quad\text{for some }2\leq k^*\leq n-1.
\end{equation}
 In  (\ref{test}), under the null, the mean vectors are constant over time while under the alternative,
there is one change-point at unknown time point $k^*$. 

Let $S(X)=X/\|X\|\mathbf{1}(X\neq \mathbf{0})$ denote the spatial sign of a vector $X$. Consider the following spatial signed SN test statistic:

\begin{equation}\label{teststat}
T_n^{(s)}:=\sup_{k=4,\cdots,n-4}\frac{(D^{(s)}(k;1,n))^2}{W_n^{(s)}(k;1,n)},
\end{equation}
where for $1\leq l\leq k<m\leq n$,
\begin{flalign}\label{DW}
D^{(s)}(k;l,m)=&\sum_{\substack{l\leq j_1,j_3\leq k\\j_1\neq j_3}}\sum_{\substack{k+1\leq j_2,j_4\leq m\\j_2\neq j_4}}S(Y_{j_1}-Y_{j_2})'S(Y_{j_3}-Y_{j_4}),\\
\label{W}W_n^{(s)}(k;l,m)=&\frac{1}{n}\sum_{t=l+1}^{k-2}D^{(s)}(t;l,k)^2+\frac{1}{n}\sum_{t=k+2}^{m-2}D^{(s)}(t;k+1,m)^2.
\end{flalign}
Here, the superscript $^{(s)}$ is used to highlight the role of spatial sign plays in constructing the testing statistic.  In contrast to \cite{wang2021inference}, where they introduced the test statistic
\begin{equation}
	T_n:=\sup_{k=2,\cdots,n-3}\frac{(D(k;1,n))^2}{W_n(k;1,n)},
\end{equation}
with $D(k;l,m)$ and $W_n(k;l,m)$  defined in the same way as (\ref{DW}) but without spatial sign.

As pointed out by \cite{wang2021inference}, the limiting distribution of (properly standardized) $D(k;1,n)$ relies heavily on the covariance (correlation) structure of $Y_i$, which is typically unknown in practice.  One may   replace it with a  consistent estimator, and this is indeed adopted in  high dimensional one-sample or two-sample testing problems, see, for example,  \cite{chen2010two} and \cite{chakraborty2017tests}. Unfortunately, in the context of change-point testing, the unknown location $k^*$ makes this method practically unreliable. For our spatial sign based test, the limiting distribution of  (properly standardized) $D^{(s)}(k;1,n)$ depends on unknown nuisance parameters that could be a complex functional of the underlying distribution.  
To this end, following \cite{wang2021inference} and \cite{zhang2021adaptive}, we propose to adopt the SN technique in \cite{shao2010testing} to avoid the consistent  estimation of unknown nuisance parameters. SN technique was initially developed  in \cite{shao2010self} and \cite{shao2010testing} in the low dimensional time series setting and its main idea is to use an inconsistent variance estimator (i.e. self-normalizer) which is based on recursive subsample test statistic, so that the limiting  distribution is pivotal under the null.  See \cite{shao2015self} for a recent review.

\section{Theoretical Properties}\label{sec:theory}
We first introduce the concept of $\rho$-mixing, see e.g. \cite{bradley1986basic}. Typical  $\rho$-mixing sequences include i.i.d sequences, $m$-dependent sequences, stationary strong ARMA processes and many Markov chain models. 

\begin{defn}[$\rho$-mixing]
A sequence $(\xi_1,\xi_2,\cdots)$ is said to be $\rho$-mixing if 
$$
\rho(d)=\sup_{k\geq 1}\sup_{f\in\mathcal{F}_1^{k},g\in\mathcal{F}_{d+k}^{\infty}}|\mathrm{Corr}(f,g)|\to 0,\quad\text{as }d\to\infty.
$$
where $\mathrm{Corr}(f,g)$ denotes the correlation between $f$ and $g$, and   $\mathcal{F}_{i}^{j}$ is the $\sigma$-field generated by $(\xi_{i},\xi_{i+1},\cdots,\xi_j)$.  Here $\rho(\cdot)$ is called the $\rho$-mixing coefficient of $(\xi_1,\xi_2,\cdots)$.
\end{defn}

\subsection{Assumptions}

To analyze the asymptotic behavior of $T_n^{(s)}$,  we make the following assumptions.
\begin{ass}\label{ass_model}
 $\{X_i\}_{i=1}^n$ are i.i.d copies of $\xi$, where $\xi$ is formed by the first $p$ observations from a sequence of  strictly stationary and $\rho$-mixing random variables $(\xi_1,\xi_2,\cdots)$ such that $E\xi_1=0$ and  $E\xi_1^2=\sigma^2$.
\end{ass}
\begin{ass}\label{ass_mixing}
	The $\rho$-mixing coefficients of $\xi$ satisfies $\sum_{k=1}^{\infty}\rho(2^k)<\infty$.
\end{ass}
Assumptions \ref{ass_model} and \ref{ass_mixing} are   imposed in \cite{chakraborty2017tests} to analyze the behaviour of spatial sign based  two-sample test statistic  for the equality of high dimensional mean. In particular,  Assumption \ref{ass_model}   allows us to analyze the behavior of $T_n^{(s)}$ under the fixed-$n$ scenario by letting $p$ go to infinity alone. Assumption \ref{ass_mixing} allows weak dependence  among the $p$ coordinates of the data, and similar assumptions are also made in, e.g. \cite{wang2021inference} and \cite{zhang2021adaptive}.  The strict stationary assumption  can be relaxed with additional conditions and the scenario that corresponds to strong coordinate-wise dependence is provided in Section \ref{sec:rsrm}

\subsection{Limiting Null}
We begin by deriving the limiting distribution of $T_n^{(s)}$ when $n$ is fixed while letting $p\to\infty$, and then analyze the large sample behavior of the fixed-$n$ limit by letting $n\to\infty$. The sequential asymptotics is fairly common in statistics and econometrics, see \cite{phillips1999}.

\begin{thm}\label{thm_main}
Suppose Assumption \ref{ass_model} and \ref{ass_mixing} hold, then under $H_0$:

(i) for any fixed $n\geq 8$, as $p\to\infty$, we have 
$$
T_n^{(s)}\overset{\mathcal{D}}{\rightarrow} \mathcal{T}_n,\quad\text{and}\quad T_n\overset{\mathcal{D}}{\rightarrow} \mathcal{T}_n,
$$
where 
$$
\mathcal{T}_n:= \sup_{k=4,\cdots,n-4}\frac{nG_n^2(\frac{k}{n};\frac{1}{n},1)}{\sum_{t=2}^{k-2}G_n^2(\frac{t}{n};\frac{1}{n},\frac{k}{n})+\sum_{t=k+2}^{n-2}G_n^2(\frac{t}{n};\frac{k+1}{n},1)},$$ with
\begin{flalign*}
G_n\Big(\frac{k}{n};\frac{l}{n},\frac{m}{n}\Big)=&\frac{(m-l)}{n}\frac{(m-k-1)}{n}Q_n\Big(\frac{l}{n},\frac{k}{n}\Big)+\frac{(m-l)}{n}\frac{(k-l)}{n}Q_n\Big(\frac{k+1}{n},\frac{m}{n}\Big)\\&-\frac{(k-l)}{n}\frac{(m-k-1)}{n}Q_n\Big(\frac{l}{n},\frac{m}{n}\Big),
\end{flalign*}
and $Q_n(\cdot,\cdot)$ is a centered Gaussian process defined on $[0,1]^2$ with covariance structure given by:
\begin{flalign*}
&\mathrm{Cov}(Q_n(a_1,b_1),Q_n(a_2,b_2))\\=&n^{-2}( \lfloor nb_1\rfloor\land \lfloor nb_2\rfloor-\lfloor na_1\rfloor\lor \lfloor na_2\rfloor)(\lfloor nb_1\rfloor\land \lfloor nb_2\rfloor-\lfloor na_1\rfloor\lor\lfloor na_2\rfloor+1)\mathbf{1}(b_1\land b_2>a_1\lor a_2).
\end{flalign*}
(ii) Furthermore, if $n\to\infty$, then
 \begin{equation}\label{limitT}\mathcal{T}_n\overset{\mathcal{D}}{\rightarrow}\mathcal{T}:=\sup_{r\in (0,1)}\frac{G(r;0,1)^2}{\int_{0}^{r}G(u;0,r)^2du+\int_{r}^{1}G(u;r,1)^2du},\end{equation}
with
\begin{flalign*}
G(r;a,b)=(b-a)(b-r)Q(a,r)+(r-a)(b-a)Q(r,b)-(r-a)(b-r)Q(a,b),
\end{flalign*}
and $Q(\cdot,\cdot)$ is a centered Gaussian process defined on $[0,1]^2$ with covariance structure given by:
$$
\mathrm{Cov}(Q(a_1,b_1),Q(a_2,b_2))=(b_1\land b_2-a_1\lor a_2)^2\mathbf{1}(b_1\land b_2>a_1\lor a_2).
$$
\end{thm}

Theorem \ref{thm_main} (i) states that for each fixed $n\geq 8$, when $p\to\infty$, the limiting distribution $\mathcal{T}_n$ is a functional of Gaussian process, which is pivotal and can be easily simulated, see Table \ref{tab_cri} for tabulated quantiles with $n=10,20,30,40,50,100,200$ (based on 50,000 Monte Carlo replications).  Theorem \ref{thm_main} (ii)  indicates that $\mathcal{T}_n$   converges in distribution as $n$ diverges, which is indeed supported by Table \ref{tab_cri}. In fact, $\mathcal{T}$ is exactly the same as the  limiting null distribution obtained in \cite{wang2021inference} under the joint asymptotics when both $p$ and $n$ diverge at the same time.

Our spatial signed  SN test builds on the test by  \cite{chakraborty2017tests}, where  an estimator  $\widehat{\Sigma}$ for the covariance $\Sigma$   is necessary as indicated by Section 2.1 therein. However, if the  sample size $n$ is fixed, their estimator $\widehat{\Sigma}$ is only unbiased but not consistent.
In contrast, 
the SN technique adopted in this paper enables us to avoid such estimation, and thus makes the fixed $n$  inference feasible in practice.  It is worth noting that the test statistics $T_n^{(s)}$ and $T_n$ share the same limiting null under both fixed-$n$ asymptotics and sequential asymptotics.
 
Our test statistic is based on the spatial signs and only assumes finite second moment, which is much weaker than the sixth moment in \cite{wang2021inference} under joint asymptotics of $p$ and $n$.
The fixed-$n$ asymptotics provides a better approximation to the finite sample distribution of $T_n^{(s)}$ and $T_n$ when $n$ is small or moderate. So its corresponding critical value should be preferred than the counterparts derived under the joint asymptotics.
Thus, when data is heavy-tailed and data length is short, our test is more appealing.
 
\begin{table}[H]
\caption{Simulated  $100\gamma\%$th quantiles of $\mathcal{T}_n$}
\label{tab_cri}
\centering
\begin{tabular}{lllllll}
\hline
$n\backslash\gamma$ & 80\%    & 90\%    & 95\%    & 99\%     & 99.5\%   & 99.9\%   \\ \hline
10       & 1681.46 & 3080.03 & 5167.81 & 14334.10 & 20405.87 & 46201.88 \\
20       & 719.03  & 1124.26 & 1624.11 & 3026.24  & 3810.61  & 5899.45  \\
30       & 633.70  & 965.12  & 1350.52 & 2403.64  & 2988.75  & 4748.03  \\
40       & 609.65  & 926.45  & 1283.00 & 2292.31  & 2750.01  & 4035.71  \\
50       & 596.17  & 889.34  & 1224.98 & 2186.99  & 2624.72  & 3846.51  \\
100      & 594.54   & 881.93  & 1200.31 & 2066.37  & 2482.51  & 3638.74  \\
200      & 592.10 & 878.23  & 1195.25 & 2049.32  & 2456.71  & 3533.44  \\ \hline
\end{tabular}
\end{table}

\subsection{Power Analysis}
Denote  $\delta=\mu_n-\mu_1$ as the shift in mean under the alternative, and $\iota^2=\lim_{p\to\infty}p^{-1}\|\delta\|^2$ as the limiting average signal. Next, we study the behavior of  the test under both fixed ($\iota>0$) and local alternatives  ($\iota=0$). 

We first consider the case when the average signal is non-diminishing.
\begin{ass}\label{ass_fix}
(i) $\iota>0$, (ii) $np\|\Sigma\|_F^{-1}\to\infty$ as $p\to\infty$.
\end{ass}
Here the Assumption \ref{ass_fix} (ii) is quite mild and can be satisfied by many weak dependent sequences such as ARMA sequences.

\begin{thm}\label{thm_fix}
[Fixed Alternative] Suppose Assumptions \ref{ass_model}--\ref{ass_fix} hold, then
$$
T_n^{(s)}\to_p\infty, \quad T_n\to_p\infty$$ 
\end{thm}

Theorem \ref{thm_fix} shows that when average signal is non-diminishing, then both $T_n^{(s)}$ and $T_n$ are consistent tests. Next, we analyze $T_n^{(s)}$ under local alternatives when $\iota=0$.  

\begin{ass}\label{ass_power}
(i) $\iota=0$, (ii) $\delta'\Sigma\delta=o(\|\Sigma\|_F^2)$ as $p\to\infty$.
\end{ass}
 Assumption \ref{ass_power} regulates the behavior of the shift size, and is used to simplify the theoretical analysis of $T_n^{(s)}$ under local alternatives.  Similar assumptions are also made in \cite{chakraborty2017tests}.  
Clearly,  when $\Sigma$ is the identity matrix, Assumption \ref{ass_power} (ii) automatically holds if $\iota=0$.

\begin{figure}[!h]
    \centering
    \includegraphics[width=0.7\textwidth]{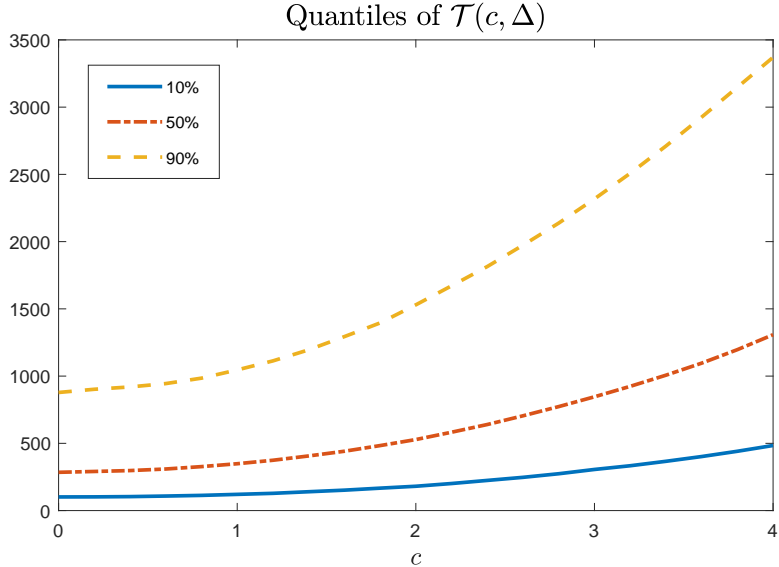}
    \caption{$\mathcal{T}(c,\Delta)$  at 10\%, 50\% and 90\% quantile levels. }
    \label{fig:local}
\end{figure}

\begin{thm}\label{thm_power}
[Local Alternative] Suppose Assumptions \ref{ass_model}, \ref{ass_mixing} and \ref{ass_power} hold. Assume there exists a $k^*$ such that $\mu_i=\mu$, $i=1\cdots, k^*$ and $\mu_i=\mu+\delta$, $i=k^*+1,\cdots,n$. Then for any fixed $n$, as $p\to\infty$,

(i) if $np\|\Sigma\|_F^{-1}\iota^2\to\infty$, then $T_n^{(s)}\to_p\infty$ and $T_n\to_p\infty$;

(ii)if $np\|\Sigma\|_F^{-1}\iota^2\to0$, then $T_n^{(s)}\overset{\mathcal{D}}{\rightarrow} \mathcal{T}_n$ and $T_n\overset{\mathcal{D}}{\rightarrow} \mathcal{T}_n$;

(iii)if $np\|\Sigma\|_F^{-1}\iota^2\to c_n\in(0,\infty)$, then 
$
T_n^{(s)}\overset{\mathcal{D}}{\rightarrow} \mathcal{T}_n(c_n,\Delta_n),
$ and $
T_n\overset{\mathcal{D}}{\rightarrow} \mathcal{T}_n(c_n,\Delta_n),
$ 
where \begin{flalign*} &\mathcal{T}_n(c_n,\Delta_n)\\=&\sup_{k=4,\cdots,n-4}\frac{n[\sqrt{2}G_n(\frac{k}{n};\frac{1}{n},1)+c_n\Delta_n(\frac{k}{n};\frac{1}{n},1)]^2}{\sum_{t=2}^{k-2}[\sqrt{2}G_n(\frac{t}{n};\frac{1}{n},\frac{k}{n})+c_n\Delta_n(\frac{t}{n};\frac{1}{n},\frac{k}{n})]^2+\sum_{t=k+2}^{n-2}[\sqrt{2}G_n(\frac{t}{n};\frac{k+1}{n},1)+c_n\Delta_n(\frac{t}{n};\frac{k+1}{n},1)]^2},\end{flalign*}and $$
\Delta_n\Big(\frac{k}{n};\frac{l}{n},\frac{m}{n}\Big)=
\begin{cases}
\frac{4{k-l+1\choose 2}{m-k^*\choose 2}}{n^4},&l<k\leq k^*<m;\\
\frac{4{k^*-l+1\choose 2}{m-k\choose 2}}{n^4},&l<k^*<k<m;\\
0,&\text{otherwise}.
\end{cases}
$$
Furthermore, if $\lim_{n\to\infty}c_n=c\in(0,\infty)$, then as $n\to\infty$,
\begin{equation}\label{TND}
    \mathcal{T}_n(c_n,\Delta_n)\overset{\mathcal{D}}{\rightarrow}  \mathcal{T}(c,\Delta)
\end{equation}
where $$
\mathcal{T}(c,\Delta):=\sup _{r \in[0,1]} \frac{\{\sqrt{2} G(r ; 0,1)+c \Delta(r, 0,1)\}^{2}}{\int_{0}^{r}\{\sqrt{2} G(u ; 0, r)+c \Delta(u, 0, r)\}^{2} d u+\int_{r}^{1}\{\sqrt{2} G(u ; r, 1)+c \Delta(u, r, 1)\}^{2} d u},
$$
 and for $b^*=\lim_{n\to\infty}(k^*/n)$,
$$
\Delta(r, a, b):= \begin{cases}\left(b^{*}-a\right)^{2}(b-r)^{2}, & a<b^{*} \leq r<b; \\ (r-a)^{2}\left(b-b^{*}\right)^{2}, & a<r<b^{*}<b; \\ 0, & \text{otherwise}.\end{cases}
$$
\end{thm}
The above theorem implies that the asymptotic power of $T_n^{(s)}$ and $T_n$ depends on the joint behavior of $\delta$ and $\|\Sigma\|_F$, holding $n$ as fixed. If $\Sigma$ is the identity matrix, then  $T_n^{(s)}$ and $T_n$ will exhibit different power behaviors according to whether $\|\delta\|/p^{1/4}$ converges to zero, infinity, or some constant $c_n\in(0,\infty)$.  
In addition, under the local alternative, the limiting distribution of $T_n^{(s)}$ and $T_n$ under the sequantial asymptotics  coincides with that  in \cite{wang2021inference} under the joint asymptotics, see Theorem 3.5 therein.   In Figure \ref{fig:local}, we plot  $\mathcal{T}(c,\Delta)$ at 10\%, 50\% and 90\% quantile levels with $b^*$  fixed  at $1/2$ and it suggests that  $\mathcal{T}(c,\Delta)$ is stochastically increasing with $c$, which further supports the consistency of both tests. 

\subsection{Analysis Under Stronger Dependence Structure}\label{sec:rsrm}
In this section, we focus on a special class  of probability models for high dimensional data termed ``randomly scaled $\rho$-mixing  (RSRM)" sequence.
\begin{defn}[RSRM, \cite{chakraborty2017tests}]
 A sequence $(\eta_1,\eta_2,\cdots)$ is a randomly scaled $\rho$-mixing sequence if there exist a zero mean $\rho$-mixing squence $(\xi_1,\xi_2,\cdots)$ and an independent positive non-degenerate random variable $R$ such that $\eta_i=\xi_i/R$, $i=1,2,\cdots$.
\end{defn}
RSRM sequences  introduce stronger dependence structure among the coordinates than $\rho$-mixing sequences, and many models fall into this category, see, e.g. elliptically symmetric models in  \cite{wang2015high} and non-Gaussian sequences in \cite{cai2014two}.

\begin{ass}\label{ass_RSRM}
Suppose $Y_i=X_i/R_i+\mu_i$,  where $\{X_i\}_{i=1}^n$ satisfies Assumptions \ref{ass_model} and \ref{ass_mixing}, and $\{R_i\}_{i=1}^n$ are i.i.d. copies of a positive random variable $R$.
\end{ass}
Clearly, when $R$ is degenerate (i.e., a positive constant), Assumption \ref{ass_RSRM} reduces to the model assumed in previous sections. However, when $R$ is non-degenerate,  Assumption \ref{ass_RSRM} imposes stronger dependence structure on coordinates of $Y_i$ than $\rho$-mixing sequences, and hence  result in additional   theoretical difficulties.  We refer to \cite{chakraborty2017tests} for more discussions of RSRM sequences.
	
	

\begin{thm}\label{thm_rsrm}
    Suppose Assumption \ref{ass_RSRM} holds, then under $H_0$,  
    
	(i) let $\mathcal{R}_n=\{R_i\}_{i=1}^n,$ for any fixed $n\geq 8$, if $\mathbb{E}(R_i^2)<\infty$ and $\mathbb{E}(R_i^{-2})<\infty$, as $p\to\infty$, there exists two random variables $\mathcal{T}_n^{(\mathcal{R}_n,s)}$ and $\mathcal{T}_n^{(\mathcal{R}_n)}$ dependent on $\mathcal{R}_n$ such that, 
	$$
	T_n^{(s)}\overset{\mathcal{D}}{\rightarrow}\mathcal{T}_n^{(\mathcal{R}_n,s)},\quad T_n\overset{\mathcal{D}}{\rightarrow}\mathcal{T}_n^{(\mathcal{R}_n)}.$$
	

	(ii)
	Furthermore, if we further assume $\mathbb{E}(R_i^4)<\infty$ and $\mathbb{E}(R_i^{-4})<\infty$, then as $n\to\infty$, we have $$\mathcal{T}_n^{(\mathcal{R}_n,s)} \overset{\mathcal{D}}{\rightarrow} \mathcal{T},\quad \mathcal{T}_n^{(\mathcal{R}_n)} \overset{\mathcal{D}}{\rightarrow} \mathcal{T}, $$
	where $\mathcal{T}$ is defined in (\ref{limitT}).
	
\end{thm}

In general, if the sample size $n$ is small and $Y_i$ is generated from an RSRM sequence,  the unconditional limiting distributions of $T_n^{(s)}$ and $T_n$ as $p\to\infty$  are no longer   pivotal  due to the randomness in $R_i$. Nevertheless, using the pivotal limiting distribution $\mathcal{T}_n$ in  hypothesis testing can still deliver relatively good performance for $T_n^{(s)}$ in both size and power, see Section \ref{sec:size} below for numerical evidence. 
If $n$ is also diverging, the same pivotal limiting distribution as presented  in Theorem \ref{thm_main} (ii) and in  Theorem 3.4 of \cite{wang2021inference} can still be reached.   

Let $\Sigma_Y$ be the covariance of $Y_i$ (or equivalently $X_i/R_i$), the next theorem provides with the asymptotic behavior  under local alternative for the RSRM model. 
\begin{thm}\label{thm_rsrmpower}
Suppose Assumptions  \ref{ass_power} and  \ref{ass_RSRM} hold, then under the local alternative such that  $ n\|\Sigma_Y\|_F^{-1}\|\delta\|^2\to c_n\in (0,\infty)$, 

(i) let $\mathcal{R}_n=\{R_i\}_{i=1}^n,$ for any fixed $n\geq 8$, if $\mathbb{E}(R_i^2)<\infty$ and $\mathbb{E}(R_i^{-2})<\infty$, as $p\to\infty$, there exists two random variables $\mathcal{T}_n^{(\mathcal{R}_n,s)}(\Delta_n^{(\mathcal{R}_n,s)})$ and $\mathcal{T}_n^{(\mathcal{R}_n)}(\Delta_n)$ dependent on $\mathcal{R}_n$ such that,  $$T_n^{(s)}\overset{\mathcal{D}}{\rightarrow}\mathcal{T}_n^{(\mathcal{R}_n,s)}(c_n,\Delta_n^{(\mathcal{R}_n,s)}),\quad T_n\overset{\mathcal{D}}{\rightarrow}\mathcal{T}_n^{(\mathcal{R}_n)}(c_n,\Delta_n).$$
(ii)
	Furthermore, if we assume $\mathbb{E}(R_i^4)<\infty$ and $\mathbb{E}(R_i^{-4})<\infty$, and $\lim_{n\to\infty}c_n=c\in(0,\infty)$, then as $n\to\infty$, we have $$\mathcal{T}_n^{(\mathcal{R}_n,s)}(c_n,\Delta_n^{(\mathcal{R}_n,s)}) \overset{\mathcal{D}}{\rightarrow}  \mathcal{T}(Kc, \Delta),\quad \mathcal{T}_n^{(\mathcal{R}_n)}(c_n,\Delta_n^{(\mathcal{R}_n)}) \overset{\mathcal{D}}{\rightarrow}  \mathcal{T}(c, \Delta),$$
	where $\mathcal{T}(c,\Delta)$ is defined in (\ref{TND}), and $$K=\mathbb{E}^{-1}\Big[\frac{R_1R_2}{\sqrt{(R_1^2+R_3^2)(R_2^2+R_3^2)}}\Big]\mathbb{E}(R_1^{-2})\mathbb{E}^2\Big[\frac{R_1R_2}{\sqrt{R_1^{2}+R_2^2}}\Big]>1 $$
	is a constant. 
\end{thm}
For the RSRM model, similar to Theorem \ref{thm_rsrm} (i), the fixed-$n$ limiting distributions of $T_n^{(s)}$ and $T_n$   are  non-pivotal under local alternatives.  However, the distribution of $T_n^{(s)}$ under sequential limit is pivotal $\mathcal{T}(Kc, \Delta)$ while that of $T_n$  is   $\mathcal{T}(c,\Delta)$. The  multiplicative  constant $K>1$
suggests that for the RSRM model, using $T_n^{(s)}$ could be more powerful as $\mathcal{T}(c,\Delta)$ is expected to be monotone in $c$, see Figure \ref{fig:local} above. This finding coincides with \cite{chakraborty2017tests} where they showed that using spatial sign based U-statistics for testing the equality of two   high dimensional means could be more powerful than the conventional mean-based ones   in \cite{chen2010two}. Thus, when strong coordinate-wise dependence is exhibited in the data, $T_n^{(s)}$ is more preferable.

\section{Multiple Change-point Estimation}\label{sec:est}
In real applications, in addition to change-point testing, another important task is to estimate the number and locations of these change-points. 
In this section, we assume there are $m\geq 1$ change-points and are denoted by ${\bm{k}}=(k_1,k_2,\cdots,k_m)\subset \{1,2,\cdots,n\}$.
A commonly used algorithm for many practitioners would be binary segmentation (BS), where the data segments are recursively split at the maximal points of the test statistics until the null of no change-points is not rejected for each segment.  However, as criticized by many researchers, BS tends to miss potential change-points  when non-monotonic change patterns are exhibited. Hence, many algorithms have been proposed to overcome this drawback.
Among them, wild binary segmentation (WBS) by \cite{fryzlewicz2014wild} and its variants  have become increasingly popular because of their easy-to-implement procedures.  The main idea of WBS is to perform BS on randomly generated  sub-intervals  so that some sub-intervals can localize at most one change-point (with high probability). 
As pointed out by \cite{kovacs2020seeded}, WBS relies on randomly generated sub-intervals and different researchers may obtain different estimates.
Hence, \cite{kovacs2020seeded} propose seeded binary segmentation (SBS) algorithm  based on deterministic construction  of these sub-intervals with relatively cheaper computational costs so that results are replicable.  
To this end, we combine the spatial signed SN test with SBS to achieve the task of multiple change-point estimation, and we call it SBS-SN$^{(s)}$.  We first introduce the concept of seeded sub-intervals. 
\begin{defn}[Seeded Sub-Intervals, \cite{kovacs2020seeded}]\label{def:seed}
Let $\alpha\in[1/2,1)$ denote a given decay parameter. For $1\leq k \leq \lfloor\log_{1/\alpha}(n)\rfloor$ (i.e. logarithm with base $1/\alpha$) define the $k$-th layer as the collection of $n_k$ intervals of initial length $l_k$ that are
evenly shifted by the deterministic shift $s_k$ as follows:
$$
\mathcal{I}_{k}=\bigcup_{i=1}^{n_{k}}\left\{\left(\left\lfloor(i-1) s_{k}\right\rfloor,\left\lceil(i-1) s_{k}+l_{k}\right\rceil\right)\right\}
$$
where $n_{k}=2\left\lceil(1 / \alpha)^{k-1}\right\rceil-1, l_{k}=10\lceil n \alpha^{k-1}/10\rceil$ and $s_{k}=\left(n-l_{k}\right) /\left(n_{k}-1\right) .$ The overall collection of seeded intervals is
$$
\mathcal{I}_{\alpha}(n)=\bigcup_{k=1}^{\left\lceil\log _{1 / \alpha}(n)\right\rceil} \mathcal{I}_{k}.
$$
\end{defn}

Let  $\alpha\in[1/2,1)$ be a decay parameter, denote $\mathcal{I}_{\alpha}(n)$ as the set of  seeded intervals based on Definition \ref{def:seed}. 
For each sub-interval $(a,b) \in \mathcal{I}_{\alpha}(n)$, we calculate the spatial signed SN test
$$
T^{(s)}(a,b)=\max_{k\in\{ a+3,\cdots,b-4\}}\frac{(D^{(s)}(k;a,b))^2}{W_{b-a+1}^{(s)}(k;a,b)},\quad b-a\geq 7,
$$
where $D^{(s)}(k;a,b))$ and $W_{b-a+1}^{(s)}(k;a,b)$ are defined in (\ref{DW}) and (\ref{W}). We obtain the  p-value of the sub-interval test statistic $T^{(s)}(a,b)$ based on the fixed-$n$ asymptotic distribution $\mathcal{T}_{b-a+1}$.  SBS-SN$^{(s)}$ then finds the smallest p-value evaluated at all sub-intervals and compare it with  a predetermined threshold  level $\zeta_p$.  If the smallest p-value is also smaller than $\zeta_p$, denote the corresponding sub-interval where the smallest p-value is achived as $(a^*,b^*)$ and  estimate the change-point by $\hat{k}=\arg\max_{k\in\{ a^*+3,\cdots,b^*-4\}}\frac{(D^{(s)}(k;a^*,b^*))^2}{W_{b^*-a^*+1}^{(s)}(k;a^*,b^*)}$. Once a change-point is identified, SBS-SN$^{(s)}$ then divides the data sample into two subsamples accordingly and apply the same procedure to each of them. The process is implemented recursively until no change-point is detected. Details are provided in Algorithm \ref{alg}.

\begin{algorithm}[!h]
	\caption{SBS-SN$^{(s)}$}\label{alg}
	\KwIn{Data $\{Y_t\}_{t=1}^{n}$, threshold p-value $\zeta_p\in(0,1)$, SBS intervals $\mathcal{I}_{\alpha}(n)$.}
	\KwOut{Estimated number of change-points $\widehat{m}$ and estimated change-points set $\hat{\bm{k}}$}
	\KwIni{SBS-SN$^{(s)}$ $(1,n,\zeta_p)$}
	\KwPro{SBS-SN$^{(s)}$ $(a,b,\zeta_p)$}
	\eIf{$b-a+1<8$}{Stop}{$\mathcal{M}_{(a,b)}:=\{i:[a_i,b_i]\in \mathcal{I}_{\alpha}(n),[a_i,b_i]\subset[a,b], b_i-a_i+1\geq 8\}$ \;
	for each $i\in \mathcal{M}_{(a,b)}$, find the p-value $p_i$ of $T^{(s)}(a_i,b_i)$ based on $\mathcal{T}_{b_i-a_i+1}$\;
	$i^*=\arg\min_{i\in\mathcal{M}_{(a,b)}}p_i$\;
	\eIf{$p_{i^*}<\zeta_p$}{
	$k^*=\arg\max_{k\in\{ a_{i^*}+3,\cdots,b_{i^*}-4\}}\frac{(D^{(s)}(k;a_{i^*},b_{i^*}))^2}{W_{b_{i^*}-a_{i^*}+1}^{(s)}(k;a_{i^*},b_{i^*})}$	\;
	$\widehat{\bm{k}}=\widehat{\bm{k}}\cup k^*$, $\widehat{m}=\widehat{m}+1$\;
	SBS-SN$^{(s)}$ $(a,k^*,\zeta_p)$\;
	SBS-SN$^{(s)}$ $(k^*+1,b,\zeta_p)$\;
	}{Stop}}
\end{algorithm}

Our SBS-SN$^{(s)}$ algorithm differs from WBS-SN algorithm in  \cite{wang2021inference} and \cite{zhang2021adaptive} in two aspects.  First, WBS-SN is built on WBS, which relies on randomly generated intervals while SBS relies on deterministic intervals. As  documented in \cite{kovacs2020seeded},  WBS is computationally more demanding than SBS. Second, the threshold used in WBS-SN is universal for each sub-interval, depends on the sample size $n$ and dimension $p$ and needs to be simulated via extensive Monte Carlo simulations. Generally speaking, WBS-SN requires simulating a new threshold each time for a new dataset.   By contrast, our estimation procedure is based on p-values under the fixed-$n$ asymptotics, which takes into account the interval length $b-a+1$ for each sub-interval $(a,b)$. When implementing either WBS or SBS,  inevitably, there will be intervals of small lengths. Hence, the universal  threshold may not be suitable as it does not take into account the effect of different interval lengths. In order to alleviate the problem of multiple testing, we may set a small  threshold number for $\zeta_p$, such as 0.0001 or 0.0005. Furthermore, the WBS-SN requires to specify a minimal interval lentgh which can affect the finite sample performance.   In this work, when generating seed sub-intervals as in Definition \ref{def:seed}, the lengths of these intervals are set as integer values times 10 to reduce the computational cost for simulating fixed-$n$ asymptotic distribution $\mathcal{T}_n$. Therefore, we only require the knowledge of  $\{\mathcal{T}_n\}_{n=10,20,\cdots}$ for SBS-SN$^{(s)}$ to work, which can be    simulated once for good and do not change with a new dataset.

\section{Numerical Experiments}\label{sec:num}
This section examines the finite sample behavior of the proposed tests and multiple change-point  estimation  algorithm SBS-SN$^{(s)}$ via simulation studies.

\subsection{Size and Power}\label{sec:size}
We first  assess the performance of $T_n^{(s)}$ with respect to various covariance structure of the data.  Consider the following data generating process with   $p=100$ and  $n\in\{10,20,50,100,200\}$:
$$
Y_i=\delta\mathbf{1}(i>0.5n)+X_i,
$$
where $\delta$ represents the mean shift vector, and $\{X_i\}_{i=1}^n$ are i.i.d copies of $X$ based on the following specifications:
\begin{enumerate}[(i)]
\item  $X\sim \mathcal{N}(\mathbf{0},I_p)$;
\item $X\sim t_5(I_p)$;
\item $X\sim t_3(I_p)$;
\item $X=(X^{(1)},\cdots,X^{(p)})'$, $X^{(t)}=\rho X^{(t-1)}+\epsilon_t$, $t=1,\cdots,p$, where $\epsilon_t\sim \mathcal{N}(0,1)/2$ are i.i.d random variables;
\item $X=(X^{(1)},\cdots,X^{(p)})'$, $X^{(t)}=\rho X^{(t-1)}+\epsilon_t$, $t=1,\cdots,p$, where $\epsilon_t\sim t_5/2$ are i.i.d random variables;
\item $X=R/U$, $R=(R^{(1)},\cdots,R^{(p)})'$, $R^{(t)}=\rho R^{(t-1)}+\epsilon_t$, $t=1,\cdots,p$, where $\epsilon_t\sim \mathcal{N}(0,1)/2$ are i.i.d random variables, and $U\sim \text{Exp}(1)$ is independently generated;
\item $X=R/U$, $R=(R^{(1)},\cdots,R^{(p)})'$, $R^{(t)}=\rho R^{(t-1)}+\epsilon_t$, $t=1,\cdots,p$, where $\epsilon_t\sim t_5/2$ are i.i.d random variables, and $U\sim \text{Exp}(1)$ is independently generated;
\end{enumerate}
where $t_{\nu}(I_p)$ is the multivariate $t$ distribution with degree of freedom $\nu$ and covariance $I_p$; $\text{Exp}(1)$  is the exponential distribution with mean $1$.

Case (i) assumes that coordinates of $X$ are independent and light-tailed; Cases (ii) and (iii) consider  the scenario of heavy-tailedness of $X$; Cases (iv) and (v) assume the  coordinates of $X$ are consecutive random observations from a stationary AR(1) model with autoregressive coefficient $\rho=0.7$; and Cases (vi) and (vii) assume the coordinates of $X$ are generated from an RSRM   with $\rho=0.7$.

Table \ref{tab_simu} shows the empirical rejection rate of $T_n$ and $T_n^{(s)}$ in percentage based on 1000 replications under the null with $H_0:$ $\delta=\mathbf{0}$; dense alternative $H_a^{1}: \delta=1/\sqrt{p}\mathbf{1}_p$; and  sparse alternative $H_a^{2}:\delta=(\mathbf{1}_2^{\top},\mathbf{0}_{p-2}^{\top})^{\top}$.   We compare the approximation using the limiting null distribution of fixed-$n$ asymptotics $\mathcal{T}_n$ and  sequential asymptotics $\mathcal{T}$ at $5\%$ level.

\begin{table}[H]
\centering
\caption{Size and power comparison of $T_n$ and $T_n^{(s)}$}
\resizebox{0.9\textwidth}{!}{
\label{tab_simu}
	\renewcommand\tabcolsep{2pt}
\centering
\begin{tabular}{ccccccccccccccccccccc}
\hline
                       &                              &                 &     & \multicolumn{5}{c}{$H_0$}        &  & \multicolumn{5}{c}{$H_a^1$}      &  & \multicolumn{5}{c}{$H_a^2$}      \\ \cline{5-9} \cline{11-15} \cline{17-21} 
Case                   & Test                         & Limit           & $n$ & 10   & 20   & 50   & 100  & 200  &  & 10   & 20   & 50   & 100  & 200  &  & 10   & 20   & 50   & 100  & 200  \\ \cline{1-4} \cline{5-9} \cline{11-15} \cline{17-21} 
\multirow{4}{*}{(i)}   & \multirow{2}{*}{$T_n$}       & $\mathcal{T}_n$ &     & 5.6  & 4.8  & 6.9  & 4.0  & 6.4  &  & 6.3  & 6.5  & 15.2 & 34.7 & 77.1 &  & 7.3  & 11.0 & 34.1 & 78.0 & 99.9 \\
                       &                              & $\mathcal{T}$   &     & 27.4 & 9.0  & 7.4  & 4.1  & 6.7  &  & 29.9 & 11.2 & 16.5 & 35.2 & 77.8 &  & 33.7 & 18.1 & 34.9 & 78.1 & 99.9 \\
                       & \multirow{2}{*}{$T_n^{(s)}$} & $\mathcal{T}_n$ &     & 5.5  & 4.8  & 6.2  & 4.3  & 6.6  &  & 6.2  & 5.9  & 15.0 & 33.4 & 76.7 &  & 7.2  & 10.4 & 33.3 & 77.7 & 99.8 \\
                       &                              & $\mathcal{T}$   &     & 28.5 & 8.7  & 6.8  & 4.4  & 7.1  &  & 29.8 & 10.8 & 15.7 & 34.6 & 77.6 &  & 33.5 & 17.3 & 34.6 & 78.6 & 99.8 \\ \hline
\multirow{4}{*}{(ii)}  & \multirow{2}{*}{$T_n$}       & $\mathcal{T}_n$ &     & 6.9  & 6.4  & 6.8  & 4.3  & 6.0  &  & 7.2  & 7.2  & 11.7 & 18.5 & 41.4 &  & 8.0  & 8.8  & 22.0 & 47.2 & 87.3 \\
                       &                              & $\mathcal{T}$   &     & 31.8 & 12.6 & 7.6  & 4.3  & 6.2  &  & 31.8 & 12.4 & 12.8 & 19.0 & 42.5 &  & 33.9 & 15.3 & 22.9 & 47.5 & 87.4 \\
                       & \multirow{2}{*}{$T_n^{(s)}$} & $\mathcal{T}_n$ &     & 5.3  & 5.3  & 6.2  & 4.1  & 5.6  &  & 5.7  & 5.5  & 11.8 & 26.1 & 59.6 &  & 6.4  & 8.1  & 26.7 & 62.9 & 96.8 \\
                       &                              & $\mathcal{T}$   &     & 28.2 & 9.7  & 6.7  & 4.2  & 5.7  &  & 28.5 & 10.0 & 12.6 & 27.0 & 60.4 &  & 30.8 & 14.4 & 28.0 & 63.2 & 96.8 \\ \hline
\multirow{4}{*}{(iii)} & \multirow{2}{*}{$T_n$}       & $\mathcal{T}_n$ &     & 9.0  & 9.5  & 9.2  & 6.7  & 7.9  &  & 10.0 & 10.4 & 11.8 & 14.2 & 25.7 &  & 9.8  & 12.3 & 17.9 & 27.9 & 57.5 \\
                       &                              & $\mathcal{T}$   &     & 35.8 & 16.1 & 9.6  & 6.9  & 8.5  &  & 35.6 & 16.1 & 12.6 & 14.7 & 26.0 &  & 36.8 & 18.7 & 18.8 & 28.7 & 58.2 \\
                       & \multirow{2}{*}{$T_n^{(s)}$} & $\mathcal{T}_n$ &     & 5.6  & 5.0  & 6.4  & 4.8  & 6.4  &  & 5.7  & 4.9  & 10.5 & 21.2 & 50.2 &  & 6.2  & 6.9  & 21.9 & 53.0 & 93.4 \\
                       &                              & $\mathcal{T}$   &     & 27.5 & 9.6  & 7.0  & 4.9  & 6.8  &  & 29.2 & 9.3  & 11.4 & 21.7 & 50.8 &  & 28.6 & 12.8 & 23.0 & 53.9 & 93.7 \\ \hline
\multirow{4}{*}{(iv)}  & \multirow{2}{*}{$T_n$}       & $\mathcal{T}_n$ &     & 5.9  & 4.8  & 6.1  & 6.8  & 5.4  &  & 6.5  & 8.6  & 23.5 & 46.4 & 78.8 &  & 9.3  & 14.7 & 43.2 & 83.8 & 99.6 \\
                       &                              & $\mathcal{T}$   &     & 28.1 & 9.2  & 6.7  & 6.9  & 5.7  &  & 30.9 & 13.8 & 24.6 & 47.4 & 79.1 &  & 33.9 & 20.1 & 44.4 & 84.0 & 99.6 \\
                       & \multirow{2}{*}{$T_n^{(s)}$} & $\mathcal{T}_n$ &     & 4.8  & 3.9  & 6.0  & 6.3  & 5.4  &  & 5.5  & 7.1  & 22.5 & 44.2 & 77.9 &  & 6.9  & 13.0 & 41.5 & 84.1 & 99.8 \\
                       &                              & $\mathcal{T}$   &     & 27.6 & 8.2  & 6.5  & 6.6  & 5.4  &  & 30.6 & 12.2 & 23.5 & 45.3 & 78.1 &  & 33.1 & 18.8 & 43.0 & 84.6 & 99.8 \\ \hline
\multirow{4}{*}{(v)}   & \multirow{2}{*}{$T_n$}       & $\mathcal{T}_n$ &     & 7.0  & 7.6  & 6.0  & 6.8  & 6.1  &  & 8.6  & 11.1 & 17.5 & 30.1 & 54.2 &  & 9.4  & 11.3 & 26.7 & 56.7 & 94.0 \\
                       &                              & $\mathcal{T}$   &     & 33.5 & 12.9 & 6.6  & 7.2  & 6.1  &  & 33.6 & 16.9 & 17.9 & 30.4 & 54.6 &  & 34.4 & 18.3 & 27.9 & 57.1 & 94.2 \\
                       & \multirow{2}{*}{$T_n^{(s)}$} & $\mathcal{T}_n$ &     & 5.3  & 4.4  & 5.0  & 6.9  & 5.2  &  & 6.1  & 7.5  & 18.5 & 37.1 & 65.2 &  & 5.6  & 9.4  & 35.2 & 73.7 & 98.7 \\
                       &                              & $\mathcal{T}$   &     & 29.3 & 8.5  & 5.3  & 7.5  & 5.5  &  & 30.5 & 11.8 & 19.0 & 37.7 & 65.6 &  & 30.6 & 14.1 & 35.8 & 74.2 & 98.8 \\ \hline
\multirow{4}{*}{(vi)}  & \multirow{2}{*}{$T_n$}       & $\mathcal{T}_n$ &     & 34.7 & 39.7 & 39.2 & 34.6 & 33.6 &  & 34.6 & 40.7 & 39.4 & 35.6 & 34.2 &  & 35.0 & 39.6 & 40.1 & 34.3 & 33.8 \\
                       &                              & $\mathcal{T}$   &     & 60.2 & 46.7 & 40.5 & 34.9 & 34.1 &  & 62.5 & 47.5 & 40.3 & 36.1 & 34.8 &  & 60.6 & 46.9 & 41.0 & 34.4 & 34.1 \\
                       & \multirow{2}{*}{$T_n^{(s)}$} & $\mathcal{T}_n$ &     & 5.0  & 4.2  & 5.3  & 5.9  & 5.9  &  & 6.0  & 4.8  & 11.3 & 20.1 & 35.3 &  & 5.6  & 7.1  & 16.8 & 37.2 & 73.5 \\
                       &                              & $\mathcal{T}$   &     & 27.9 & 8.6  & 5.7  & 6.2  & 6.1  &  & 28.1 & 10.0 & 12.0 & 20.3 & 35.4 &  & 28.2 & 11.9 & 17.6 & 38.0 & 74.0 \\ \hline
\multirow{4}{*}{(vii)} & \multirow{2}{*}{$T_n$}       & $\mathcal{T}_n$ &     & 33.7 & 40.6 & 37.9 & 36.5 & 36.6 &  & 34.3 & 40.3 & 37.9 & 37.0 & 36.9 &  & 33.5 & 40.6 & 38.3 & 36.9 & 36.8 \\
                       &                              & $\mathcal{T}$   &     & 61.9 & 47.3 & 38.6 & 37.2 & 36.8 &  & 62.2 & 46.5 & 39.1 & 37.4 & 37.1 &  & 61.5 & 47.7 & 39.8 & 37.7 & 36.9 \\
                       & \multirow{2}{*}{$T_n^{(s)}$} & $\mathcal{T}_n$ &     & 4.3  & 4.4  & 5.2  & 6.4  & 6.0  &  & 5.1  & 6.2  & 9.5  & 17.5 & 32.9 &  & 5.1  & 5.8  & 14.1 & 28.5 & 62.5 \\
                       &                              & $\mathcal{T}$   &     & 30.2 & 8.4  & 5.5  & 6.7  & 6.5  &  & 30.6 & 10.1 & 10.2 & 17.7 & 33.5 &  & 30.4 & 9.2  & 15.3 & 29.1 & 63.0 \\ \hline 
\end{tabular}}
\end{table}
	
We summarize the findings of Table \ref{tab_simu} as follows: (1) both $T_n$ and $T_n^{(s)}$ suffer from severe size distortion using sequential asymptotics $\mathcal{T}$ if $n$ is small (i.e., $n=10,20,50$);  (2) both  fixed-$n$ asymptotics $\mathcal{T}_n$ and  large-$n$ asymptotics $\mathcal{T}$ work well for $T_n$ and $T_n^{(s)}$ when $n$ is large under weak dependence in coordinates (cases (i)-(v)); (3) $T_n$ and $T_n^{(s)}$ are both accurate in size and comparable in power performance when $X_i$'s are light-tailed (cases (i),(ii), (iv) and (v)) if appropriate limiting distributions are used; (4) $T_n$ is slightly oversized compared with $T_n^{(s)}$ under heavy-tailed distributions (case (iii)); (5) when strong dependence is exhibited in coordinates (cases (vi) and (vii)),  $(T_n^{(s)},\mathcal{T}_n)$  still works  for small $n$ while other combinations of tests and asymptotics generally fail; (6) increasing the data length $n$ enhances power under all settings while increasing dependence in coordinates   generally reduces power. Overall, the spatial signed SN test using fixed-$n$ asymptotic critical value outperforms all other tests and should be preferred due to its robustness and size accuracy.

\subsection{Segmentation}
We then examine the numerical performance of SBS-SN$^{(s)}$ by considering the multiple change-points models in \cite{wang2021inference}. For $p=50$ and $n=120$, we generate i.i.d. samples of $\{X_i\}_{i=1}^{n}$, where $X_i$'s are either normally distributed, i.e., case (i); or RSRM sequences, i.e., case (vii) with autoregressive coefficient $\rho=0.3.$  We assume there are three change points  at $k=$ 30, 60 and 90. Denote the changes in mean by $\bm{\theta}_1$, $\bm{\theta}_2$ and $\bm{\theta}_3$, where $\bm{\theta}_1=-\bm{\theta}_2=\bm{\theta}_3=\sqrt{h/d}(\bm{1}_{d}^{\top},\bm{0}_{p-d}^{\top})^{\top}$,  $d\in\{5,p\}$ and $h\in\{2.5,4\}.$ Here, $d$ represents the sparse or dense alternatives while $h$ represents the signal strength. For example, we refer to the choice of $d=5$ and $h=2.5$ as Sparse(2.5) and $d=p$, $h=4$ as Dense(4).

To assess the estimation accuracy, we consider (1) the differences between the estimated number of change-points $\hat{m}$ and the truth $m=3$; (2) Mean Squared Error (MSE) between $\hat{m}$ and the truth $m=3$; (3) the adjusted Rand index (ARI) which measures the similarity between two partitions of the same observations. Generally speaking, a higher ARI (with the
maximum value of 1)  indicates more accurate change-point estimation, see \cite{hubert1985comparing} for more details. 

We also  report the estimation results using   WBS-SN in \cite{wang2021inference} and SBS-SN (by replacing WBS with SBS in \cite{wang2021inference}). The superiority of WBS-SN over other competing methods can be found in \cite{wang2021inference} under similar simulation settings.  For SBS based estimation results, we vary the decay rate $\alpha\in\{2^{-1/2}, 2^{-1/4}\}$, which are denoted by SBS$^{1}$ and SBS$^{2}$ respectively. Here, the thresholds in above algorithms are either  Monte
Carlo simulated constant $\zeta_n$ for all sub-interval test statistics; or prespecified quantile level  $\zeta_p\in(0,1)$ based on p-values reported by these sub-interval  statistics. Here, we fix $\zeta_n$ as the 95\% quantile levels of the maximal test statistics based on 5000 simulated Gaussian data with no change-points as in \cite{wang2021inference} while $\zeta_p$ is set as 0.001 (the results using 0.005 is similar hence omitted). We replicate all experiments 500 times, and results for dense changes and sparse changes are reported in Table \ref{tab_dense} and Table \ref{tab_sparse}, respectively. 

From Table \ref{tab_dense}, we find that (1) WBS-SN and SBS-SN tend to produce close estimation accuracy  holding the form of $X$ and signal strength fixed;  (2) different decay rates have some impact  on the performance of    SBS-SN methods, and when the signal is weak the impact is noticeable;   (3) increasing the signal strength of change-points   boosts the detection power for all methods; (4) using $\zeta_p$ gives more accurate estimation than using  $\zeta_n$ in SBS-SN when data is normally distributed with weak signal level and they work comparably in other settings; (5) when $X$ is normally distributed, SBS-SN$^{(s)}$ works comparably with other estimation algorithms while for RSRM sequences,  our SBS-SN$^{(s)}$ 
greatly outperforms other methods.  The results in Table \ref{tab_sparse} are similar.    These findings together suggest 
substantial gain in detection performance using SBS-SN$^{(s)}$ due to its robustness to heavy-tailedness and stronger dependence in the coordinates.

\begin{table}[H]
\centering
\caption{Estimation results with dense changes.}
\label{tab_dense}
\resizebox{0.9\textwidth}{!}{
\begin{tabular}{cccccccccccc}
\hline
                              &         & Test        & Threshold           & $X$    & \multicolumn{5}{c}{$\hat{m}-m$}                 & MSE   & ARI   \\ \cline{6-10}
                              &         &             &                 &        & \textless{}-1 & -1  & 0   & 1   & \textgreater{}1 &       &       \\ \hline
\multirow{14}{*}{Dense(2.5)}  & WBS     & $T_n$       & $\zeta_n$   & Normal & 95            & 156 & 246 & 3   & 0               & 1.278 & 0.729 \\
                              & SBS$^1$ & $T_n$       & $\zeta_n$   & Normal & 76            & 206 & 214 & 4   & 0               & 1.068 & 0.742 \\
                              & SBS$^2$ & $T_n$       & $\zeta_n$  & Normal & 32            & 195 & 266 & 7   & 0               & 0.660 & 0.809 \\
                              & SBS$^1$ & $T_n$       & $\zeta_p$ & Normal & 0             & 0   & $\bm{464}$ & 35  & 1               & 0.078 & $\bm{0.929}$ \\
                              & SBS$^2$ & $T_n$       & $\zeta_p$& Normal & 0             & 10  & $\bm{ 466}$ & 23  & 1               & $\bm{0.074}$ & $\bm{0.931}$ \\
                              & SBS$^1$ & $T_n^{(s)}$ & $\zeta_p$& Normal & 0             & 1   &$\bm {469}$ & 29  & 1               & $\bm{0.068}$ & $\bm{0.928}$ \\
                              & SBS$^2$ & $T_n^{(s)}$ & $\zeta_p$& Normal & 0             & 19  & 463 & 18  & 0               & $\bm { 0.074}$ & 0.926 \\
                              & WBS     & $T_n$       & $\zeta_n$   & RSRM   &64               &89     &133     &104     & 110                &2.668       &0.461       \\
                              & SBS$^1$ & $T_n$       & $\zeta_n$   & RSRM   & 35            & 58  & 116 & 107 & 184             & 3.612 & 0.462 \\
                              & SBS$^2$ & $T_n$       & $\zeta_n$   & RSRM   & 38            & 72  & 114 & 130 & 146             & 3.032 & 0.470 \\
                              & SBS$^1$ & $T_n$       & $\zeta_p$ & RSRM   & 8             & 6   & 32  & 84  & 376             & 8.792 & 0.653 \\
                              & SBS$^2$ & $T_n$       & $\zeta_p$& RSRM   & 57            & 43  & 104 & 106 & 233             & 4.508 & 0.532 \\
                              & SBS$^1$ & $T_n^{(s)}$ & $\zeta_p$& RSRM   & 3             & 65  & 412 & 20  & 0               & 0.194 & 0.880 \\
                              & SBS$^2$ & $T_n^{(s)}$ & $\zeta_p$ & RSRM   & 11            & 121 & 355 & 13  & 0               & 0.356 & 0.850 \\ \hline 
\multirow{14}{*}{Dense(4)}    & WBS     & $T_n$       & $\zeta_n$   & Normal &0             & 7  & 486 & 7  & 0               & 0.028 & 0.948 \\
                              & SBS$^1$ & $T_n$       & $\zeta_n$   & Normal & 0             & 21  & 467 & 12  & 0               & 0.066 & 0.936 \\
                              & SBS$^2$ & $T_n$       & $\zeta_n$   & Normal & 0             & 23  & 464 & 13  & 0               & 0.072 & 0.945 \\
                              & SBS$^1$ & $T_n$       & $\zeta_p$& Normal & 0             & 0   & 464 & 34  & 2               & 0.084 & 0.937 \\
                              & SBS$^2$ & $T_n$       & $\zeta_p$& Normal & 0             & 0   & 476 & 23  & 1               & 0.054 & 0.943 \\
                              & SBS$^1$ & $T_n^{(s)}$ & $\zeta_p$ & Normal & 0             & 0   & 468 & 31  & 1               & 0.070 & 0.936 \\
                              & SBS$^2$ & $T_n^{(s)}$ & $\zeta_p$ & Normal & 0             & 0   & 482 & 18  & 0               & 0.036 & 0.942 \\
                              & WBS     & $T_n$       & $\zeta_n$  & RSRM   & 64            & 89  & 133 & 104 & 110             & 2.668 & 0.461 \\
                              & SBS$^1$ & $T_n$       & $\zeta_n$  & RSRM   & 26            & 67  & 107 & 115 & 185             & 3.732 & 0.484 \\
                              & SBS$^2$ & $T_n$       & $\zeta_n$   & RSRM   & 33            & 74  & 115 & 125 & 153             & 3.146 & 0.489 \\
                              & SBS$^1$ & $T_n$       & $\zeta_p$& RSRM   & 27            & 25  & 71  & 93  & 309             & 6.604 & 0.579 \\
                              & SBS$^2$ & $T_n$       & $\zeta_p$ & RSRM   & 39            & 33  & 103 & 114 & 244             & 4.740 & 0.559 \\
                              & SBS$^1$ & $T_n^{(s)}$ & $\zeta_p$& RSRM   & 0             & 10  & 469 & 20  & 1               & 0.068 & 0.918 \\
                              & SBS$^2$ & $T_n^{(s)}$ & $\zeta_p$ & RSRM   & 0             & 35  & 451 & 14  & 0               & 0.098 & 0.911 \\ \hline
\end{tabular}}
\vspace{1ex}

     {\raggedright\small ~~ ~~Note: Top 3 methods are in bold format. \par}
\end{table}
\begin{table}[H]
\centering
\caption{Estimation results with sparse changes.}
\label{tab_sparse}	
\resizebox{0.9\textwidth}{!}{\begin{tabular}{cccccccccccc}

\hline
                              &         & Test        & Threshold         & $X$    & \multicolumn{5}{c}{$\hat{m}-m$}                 & MSE   & ARI   \\ \cline{6-10}
                              &         &             &                 &        & \textless{}-1 & -1  & 0   & 1   & \textgreater{}1 &       &       \\ \hline
\multirow{14}{*}{Sparse(2.5)} & WBS     & $T_n$       & $\zeta_n$   & Normal & 78            & 147 & 274 & 1   & 0               & 1.050 & 0.759 \\
                              & SBS$^1$ & $T_n$       & $\zeta_n$   & Normal & 59            & 214 & 223 & 4   & 0               & 0.928 & 0.760 \\
                              & SBS$^2$ & $T_n$       &$\zeta_n$   & Normal & 20            & 185 & 287 & 8   & 0               & 0.546 & 0.829 \\
                              & SBS$^1$ & $T_n$       & $\zeta_p$ & Normal & 0             & 1   & 464 & 34  & 1               & 0.078 & 0.929 \\
                              & SBS$^2$ & $T_n$       & $\zeta_p$& Normal & 0             & 13  & 465 & 21  & 1               & 0.076 & 0.930 \\
                              & SBS$^1$ & $T_n^{(s)}$ & $\zeta_p$ & Normal & 0             & 2   & 470 & 27  & 1               & 0.066 & 0.927 \\
                              & SBS$^2$ & $T_n^{(s)}$ & $\zeta_p$ & Normal & 0             & 23  & 460 & 17  & 0               & 0.080 & 0.924 \\
                              & WBS     & $T_n$       & $\zeta_n$  & RSRM   & 70            & 97  & 121 & 102 & 110             & 2.682 & 0.449 \\
                              & SBS$^1$ & $T_n$       & $\zeta_n$   & RSRM   & 38            & 51  & 110 & 124 & 177             & 3.572 & 0.460 \\
                              & SBS$^2$ & $T_n$       & $\zeta_n$   & RSRM   & 39            & 66  & 122 & 122 & 151             & 3.100 & 0.474 \\
                              & SBS$^1$ & $T_n$       & $\zeta_p$ & RSRM   & 38            & 35  & 75  & 109 & 278             & 5.800 & 0.447 \\
                              & SBS$^2$ & $T_n$       & $\zeta_p$ & RSRM   & 84            & 69  & 129 & 108 & 179             & 3.258 & 0.528 \\
                              & SBS$^1$ & $T_n^{(s)}$ & $\zeta_p$ & RSRM   & 5             & 64  & 414 & 17  & 0               & 0.212   & 0.872 \\
                              & SBS$^2$ & $T_n^{(s)}$ & $\zeta_p$ & RSRM   & 8             & 117 & 364 & 11  & 0               & 0.330 & 0.851 \\ \hline
\multirow{14}{*}{Sparse(4)}   & WBS     & $T_n$       & $\zeta_n$   & Normal & 0             & 7   & 486 & 7   & 0               & 0.028 & 0.958 \\
                              & SBS$^1$ & $T_n$       & $\zeta_n$   & Normal & 0             & 19  & 468 & 13  & 0               & 0.064 & 0.938 \\
                              & SBS$^2$ & $T_n$       & $\zeta_n$  & Normal & 0             & 27  & 458 & 15  & 0               & 0.084 & 0.945 \\
                              & SBS$^1$ & $T_n$       & $\zeta_p$ & Normal & 0             & 0   & 465 & 34  & 1               & 0.076 & 0.938 \\
                              & SBS$^2$ & $T_n$       & $\zeta_p$ & Normal & 0             & 0   & 477 & 22  & 1               & 0.052 & 0.944 \\
                              & SBS$^1$ & $T_n^{(s)}$ & $\zeta_p$ & Normal & 0             & 0   & 472 & 27  & 1               & 0.062 & 0.937 \\
                              & SBS$^2$ & $T_n^{(s)}$ & $\zeta_p$ & Normal & 0             & 0   & 481 & 19  & 0               & 0.038 & 0.943 \\
                              & WBS     & $T_n$       & $\zeta_n$  & RSRM   & 58            & 93  & 125 & 106 & 118             & 2.716 & 0.476 \\
                              & SBS$^1$ & $T_n$       & $\zeta_n$  & RSRM   & 32            & 51  & 96  & 133 & 188             & 3.840 & 0.486 \\
                              & SBS$^2$ & $T_n$       & $\zeta_n$  & RSRM   & 32            & 62  & 121 & 117 & 168             & 3.256 & 0.503 \\
                              & SBS$^1$ & $T_n$       & $\zeta_p$ & RSRM   & 27            & 25  & 76  & 97  & 300             & 6.250 & 0.574 \\
                              & SBS$^2$ & $T_n$       & $\zeta_p$ & RSRM   & 70            & 55  & 119 & 117 & 194             & 3.460 & 0.557 \\
                              & SBS$^1$ & $T_n^{(s)}$ & $\zeta_p$ & RSRM   & 0             & 8   & 472 & 20  & 0               & 0.056 & 0.914 \\
                              & SBS$^2$ & $T_n^{(s)}$ & $\zeta_p$ & RSRM   & 0             & 38  & 448 & 14  & 0               & 0.104 & 0.908 \\ \hline
\end{tabular}}
\end{table}
\section{Real Data Application}\label{sec:real}
In this section, we analyze the genomic micro-array (ACGH) dataset for 43 individuals with bladder tumor.  The ACGH data contains log intensity ratios of these individuals measured at 2215 different loci on their genome, and  copy number variations in the loci can be viewed as the change-point in the genome.  Hence change-point estimation could be helpful in determining the abnormality regions, as analyzed by \cite{wang2018high} and \cite{zhang2021adaptive}. The data is denoted by $\{Y_i\}_{i=1}^{2215}$. 

\begin{figure}[H]
		\centering 
		\begin{subfigure}{0.49\textwidth}
			\centering
			\includegraphics[width=1\textwidth]{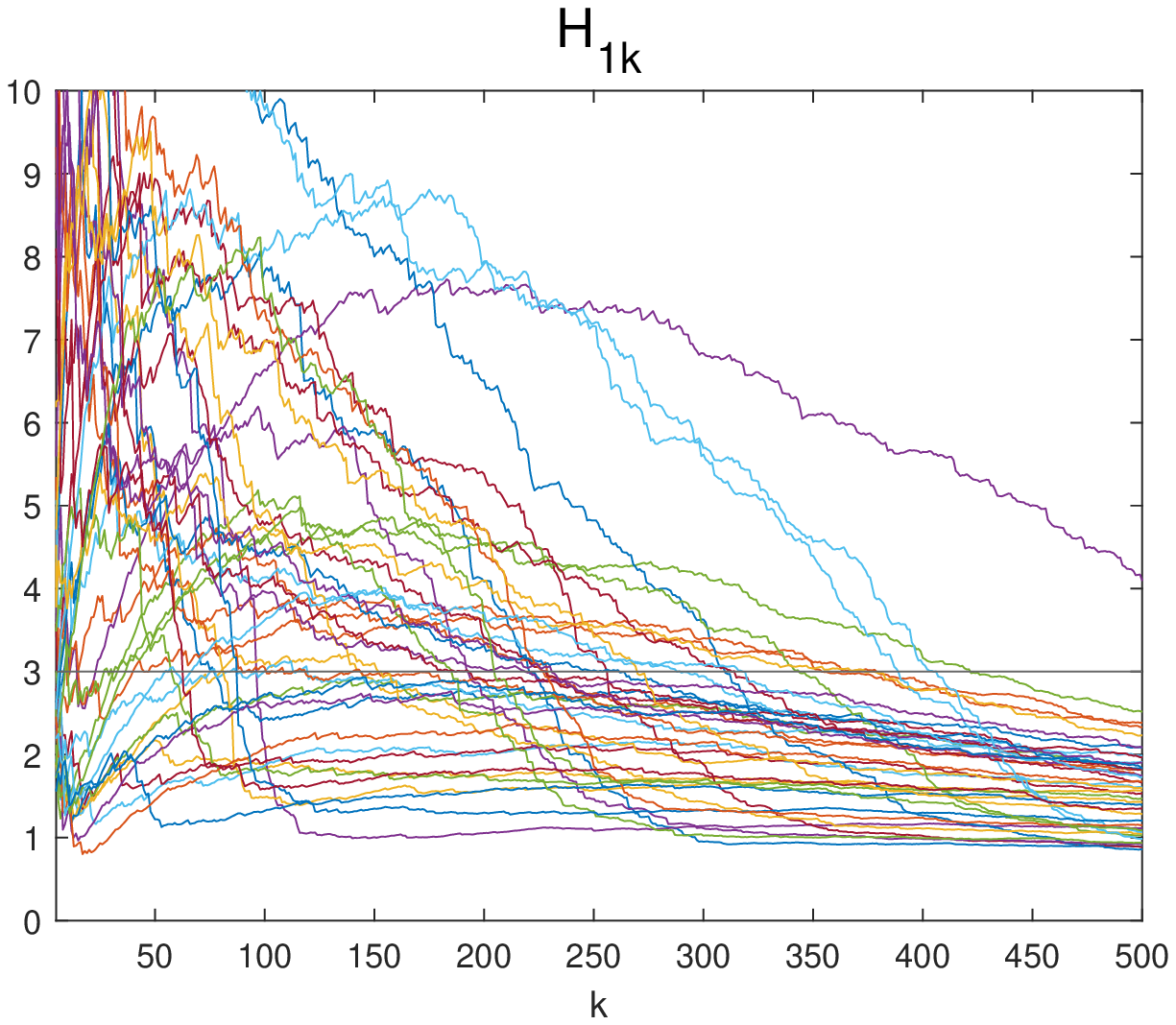}
		\end{subfigure}
		\begin{subfigure}{0.49\textwidth}
			\centering
			\includegraphics[width=1\textwidth]{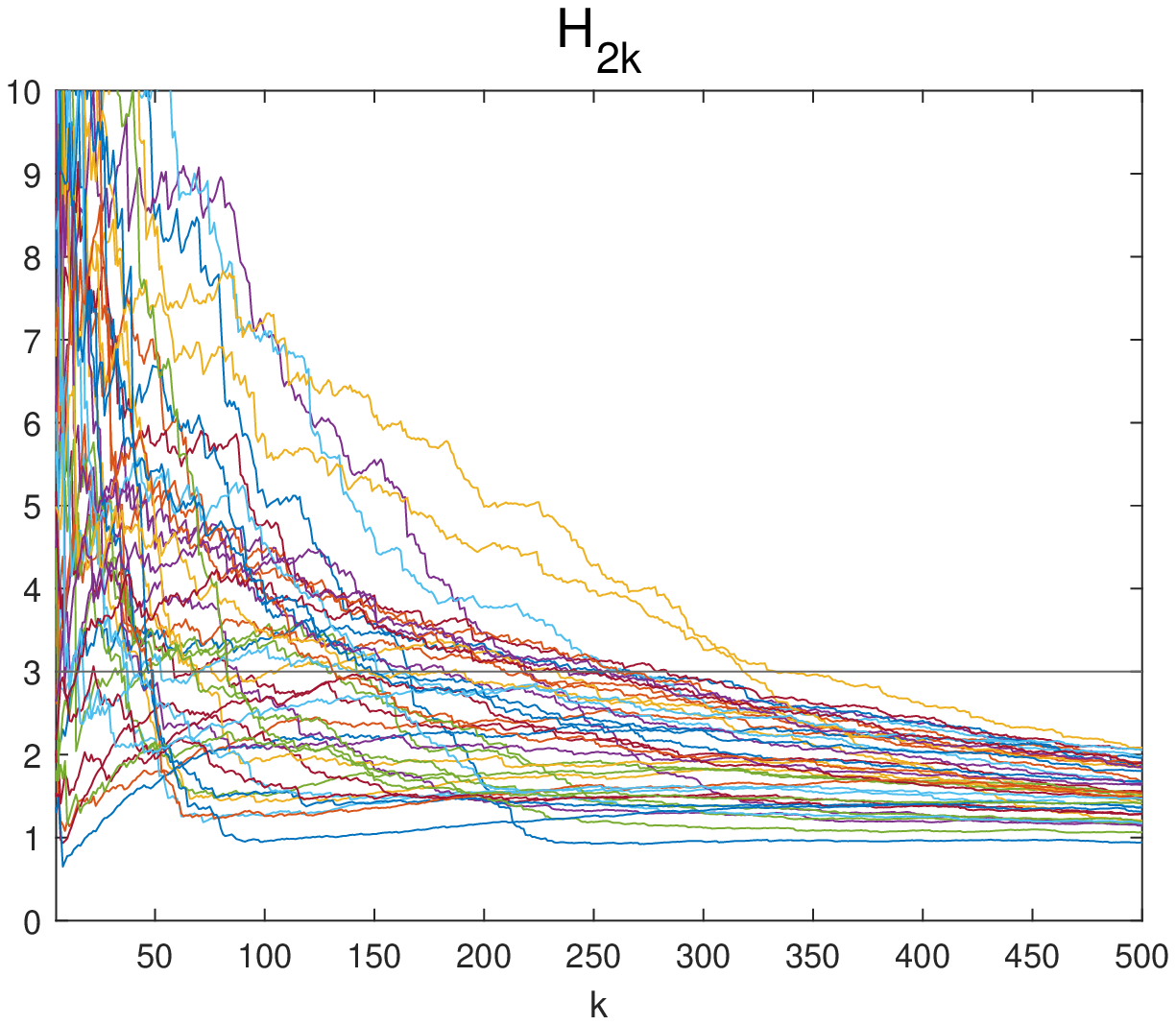}
		\end{subfigure}
		\caption{Hill's estimator for 43 individuals. }
		\label{fig:hill}
	\end{figure}

To illustrate the necessity of robust estimation method proposed in this paper,   we use the Hill's estimator to estimate the tail index of a sequence, see \cite{hill1975simple}. Specifically, let $Y_{(i),j}$ be  the ascending order statistics of the $j$th individual (coordinate) across 2215 observations. For $j=1,2,\cdots,43$, we give the left-tail and  right-tail  Hill estimators by 
$$
H_{1 k,j}=\left\{\frac{1}{k} \sum_{i=1}^{k} \log \left(\frac{Y_{(i),j}}{Y_{(k+1),j}}\right)\right\}^{-1} \quad \text { and } \quad H_{2 k,j}=\left\{\frac{1}{k} \sum_{i=1}^{k} \log \left(\frac{Y_{(n-i+1),j}}{Y_{(n-k),j}}\right)\right\}^{-1},
$$
respectively, and they are plotted in Figure \ref{fig:hill}.  From the plot, we see that 
most of the right-tail and the left-tail indices are below 3, suggesting the data is very likely heavy-tailed. 

We take the first 200 loci for our SBS-SN$^{(s)}$ change-point estimation following the practice in \cite{zhang2021adaptive}, where the decay rate for generation of seeded interval in SBS is $2^{-1/4}$.  We also compare the results obtained for Adaptive WBS-SN in \cite{zhang2021adaptive} and 20 most
significant points detected by INSPECT in \cite{wang2018high}. For this dataset, INSPECT is more like a screening method as it delivers a total of 67 change-points. In contrast to Adaptive WBS-SN and INSPECT where the thresholds for change-point estimation are simulated, the threshold used in SBS-SN$^{(s)}$ can be pre-specified, and it reflects a researcher's confidence in detecting the change-points. We set the p-value threshold $\zeta_p$  as 0.001, 0.005 and 0.01 and the results are as follows:
$$
\begin{array}{ll}
\text{Adaptive WBS-SN} & 15,32,38,44,59,74,91,97,102,116,134,158,173,186,191 \\
\text {INSPECT} & 15,26,28,33,36,40,56,73,91,97,102,119,131,134,135,146,155,\\&174,180,191\\
\text{SBS-SN$^{(s)}$, $\zeta_p=0.001$}&  30,    41,    72,    89,   130,   136,   174\\
\text{SBS-SN$^{(s)}$, $\zeta_p=0.005$}&   30,    41,    56,    72,    89,    97,   116,   130,   136,   155,   174,   191\\
\text{SBS-SN$^{(s)}$, $\zeta_p=0.01$}&   30,    41,    56,    72,    89,    97,   111,   116,   130,   136,   155,   174,   191
\end{array}
$$
As we see, increasing the p-value threshold  $\zeta_p$ leads to more estimated change-points, and the set of estimated change-points by using larger $\zeta_p$ contain those by smaller $\zeta_p$ as  subsets.   In addition, increasing $\zeta_p$ from 0.005 to 0.01 only brings in one more estimated change-point, suggesting $\zeta_p=0.005$ may be a reasonable choice for the ACGH dataset. 

 All of our detected change-points at $\zeta_p=0.005$  are also detected by INSPECT, i.e., 30(28), 41(40), 56,  72(73), 89(91), 97, 116, 130(131), 136 (134,135), 155, 174, 191. Although most of these points also coincide with Adaptive WBS-SN, there are non-overlapping ones. For example, 41, 56, 130 in SBS-SN$^{(s)}$ seem to be missed in Adaptive WBS-SN while  102 is missed by our SBS-SN$^{(s)}$ as it is detected by both   Adaptive WBS-SN and INSPECT.  These results are not really in conflict as Adaptive WBS-SN targets both sparse and dense alternatives, whereas our procedure aims to detect dense change with robustness properties.

\section{Conclusion}\label{sec:con}
In this paper, we propose a new method for testing and estimation of change-points in high dimensional independent data. Our test statistic builds   on two recent advances in high-dimensional testing problem: spatial sign used in two-sample testing in \cite{chakraborty2017tests} and self-normalized U-statistics in \cite{wang2021inference}, and inherits many advantages therein such as  robustness to heavy-tailedness and tuning-free.  
The test is theoretically justified under both fixed-$n$ asymptotics and sequential asymptotics, and under both null and alternatives. When data exhibits stronger dependence in coordinates, we further enhance the analysis by focusing on RSRM models, and discover that using spatial sign leads to power improvement compared with mean based tests in \cite{wang2021inference}.  As for   multiple change-point estimation, we  propose to combine p-values under the fixed-$n$ asymptotics with the SBS algorithm.  Numerical simulations demonstrate  that our fixed-$n$ asymptotics for spatial sign based test   provides a better approximation to the finite sample distribution, and the estimation algorithm outperforms the mean-based ones when data is heavy-tailed and when coordinates are strongly dependent.  

To conclude, we mention a few interesting topics for future research. Our method builds on spatial sign and  targets dense signals by constructing unbiased estimators for $\|\mathbb{E} S(Y_1-Y_n)\|$. As pointed out by \cite{liu2020unified}, many real data exhibit both sparse and dense changes, and it would be interesting to  combine with the adaptive SN based test in \cite{zhang2021adaptive} to achieve both robustness and adaptiveness. In addition, the independence assumption imposed in this paper may limit  its applicability to high dimensional time series where temporal dependence can not be neglected. It's desirable to relax the independence assumption by  U-statistics based on trimmed observations, as is adopted in \cite{wang2021inference}.  It would also  be interesting to develop robust methods for detecting change-points in other quantities beyond mean, such as quantiles, covariance matrices and  parameter vectors in high dimensional linear models.

\newpage

\setcounter{section}{0}
\begin{center}
	\Large Supplement to ``Robust Inference for Change Points in High Dimension"
\end{center}
\bigskip
\baselineskip=1.25\baselineskip
\renewcommand{\thetable}{S.\arabic{table}}
\renewcommand{\thefigure}{S.\arabic{figure}}
\renewcommand{\thesection}{S.\arabic{section}}
\renewcommand{\theass}{S.\arabic{ass}}
\renewcommand{\thethm}{S.\arabic{thm}}
\renewcommand{\theequation}{S.\arabic{equation}}
\spacingset{1.25}

This supplementary material contains all the technical proofs for the main paper.  Section \ref{sec:proof} contains all the proofs of main theorems and Section \ref{sec:lemma} contains auxiliary lemmas and their proofs.

In what follows, let $a_{i,k}$ denote the $k$th coordinate of a vector $a_i$. Denote $a_n\lesssim b_n$ if there exits $M,C>0$ such that $a_n\leq C b_n$ for $n>M$.

\section{Proofs of Theorems}\label{sec:proof}

\subsection{Proof of Theorem \ref{thm_main}}
First, we have that 
\begin{flalign}\label{decomp}
	\|Y_i-Y_j\|^2=\sum_{\ell=1}^{p}(X_{i,\ell}-X_{j,\ell})^2+2(\mu_j-\mu_i)'(X_{j}-X_i)+\|\mu_i-\mu_j\|^2.
\end{flalign}
(i) Under $H_0$, by Theorem 8.2.2 in \cite{zhengyan1997limit}, as $p\to\infty$, we have almost surely,
\begin{flalign}\label{consis}
	\frac{1}{p}	\|Y_i-Y_j\|^2=2\sigma^2.
\end{flalign}

Then, for any fixed $k,l,m$, we have that 
\begin{flalign}\label{D}
	\begin{split}
		&D^{(s)}(k;l,m)\\=&\sum_{\substack{l\leq j_1,j_3\leq k\\j_1\neq j_3}}\sum_{\substack{k+1\leq j_2,j_4\leq m\\j_2\neq j_4}}\frac{(Y_{j_1}-Y_{j_2})'(Y_{j_3}-Y_{j_4})}{2p\sigma^2}
		\\&+\sum_{\substack{l\leq j_1,j_3\leq k\\j_1\neq j_3}}\sum_{\substack{k+1\leq j_2,j_4\leq m\\j_2\neq j_4}}\frac{(Y_{j_1}-Y_{j_2})'(Y_{j_3}-Y_{j_4})}{2p\sigma^2}\Big\{\frac{2p\sigma^2}{\|Y_{j_1}-Y_{j_2}\|\|Y_{j_3}-Y_{j_4}\|}-1\Big\}
		\\=:&(2p\sigma^2)^{-1}[D_1(k;l,m)+D_2(k;l,m)],
	\end{split}
\end{flalign}
where clearly $D_1(k;l,m)=D(k;l,m)$, and 
\begin{flalign*}
	D_2(k;l,m)=\sum_{\substack{l\leq j_1,j_3\leq k\\j_1\neq j_3}}\sum_{\substack{k+1\leq j_2,j_4\leq m\\j_2\neq j_4}}{(Y_{j_1}-Y_{j_2})'(Y_{j_3}-Y_{j_4})}\Big\{\frac{2p\sigma^2}{\|Y_{j_1}-Y_{j_2}\|\|Y_{j_3}-Y_{j_4}\|}-1\Big\}.
\end{flalign*}

Then, Theorem 4.0.1 in \cite{zhengyan1997limit} implies that 
$$
\frac{\Gamma^{-1/2}(k;l,m)\Big\{D_1(k;l,m)\Big\}}{(m-k)(m-k-1)(k-l+1)(k-l)}\overset{\mathcal{D}}{\rightarrow} \mathcal{N}(0,1),
$$
where $\Gamma(k;l,m)=\frac{2[(m-l)(m-l-1)]}{(m-k)(m-k-1)(k-l+1)(k-l)}\mathrm{tr}(\Sigma^2)$, or equivalently 
\begin{flalign}\label{D1}
	\frac{1}{n^3\|\Sigma\|_F}\Big\{D_1(k;l,m)\Big\}\overset{\mathcal{D}}{\rightarrow} \mathcal{N}\Big(0,\frac{16{m-l\choose2}{k-l+1\choose2}{m-k\choose 2}}{n^6}\Big).
\end{flalign}

Next, since we view $n$ as fixed, then  for all $j_1\neq j_3$, $j_3\neq j_4$, by Theorem 4.0.1 in \cite{zhengyan1997limit}, it follows that $\|\Sigma\|_F^{-1}(Y_{j_1}-Y_{j_2})'(Y_{j_3}-Y_{j_4})=O_p(1)$. In addition, in view of (\ref{consis}) we have 
$\frac{2p\sigma^2}{\|Y_{j_1}-Y_{j_2}\|\|Y_{j_3}-Y_{j_4}\|}-1=o_p(1)$, and this implies that $n^{-3}\|\Sigma\|_F^{-1}D_2(k;l,m)=o_p(1)$.

Hence, combined with (\ref{D1}), we have 
\begin{flalign}\label{equiv}
	\begin{split}
		T_n^{(s)}
		=&\sup_{k=4,\cdots,n-4}\frac{\Big(2p\sigma^2 n^{-3}\|\Sigma\|_F^{-1}D^{(s)}(k;1,n)\Big)^2}{4p\sigma^4n^{-6} \|\Sigma\|_F^{-2}W_n^{(s)}(k;1,n)}
		\\=&\sup_{4,\cdots,n-4}\frac{ n^{-6}\|\Sigma\|_F^{-2}[D_1(k;1,n)+D_2(k;1,n)]^2}{n^{-6} \|\Sigma\|_F^{-2}W_{n}(k;1,n)}+o_p(1)
		\\=&T_n+o_p(1),
	\end{split}
\end{flalign}
where the last equality holds since $n^{-3}\|\Sigma\|_F^{-1}D_2(k;l,m)=o_p(1)$ for each triplet $(k,l,m)$.

For $0\leq k<m\leq n$, we let
$$
Z(k,m)=\sum_{i=k+1}^{m}\sum_{j=k}^{i-1}X_i'X_j,
$$
then it follows that 
\begin{flalign}\label{decompD}
	\begin{split}
		D(k;l,m)=&2(m-k)(m-k-1)Z(l,k)+2(k-l+1)(k-l)Z(k+1,m)\\&-2(k-l)(m-k-1)[Z(l,m)-Z(l,k)-Z(k+1,m)].
	\end{split}
\end{flalign}
Then, by Lemma \ref{lem_fix}, and continuous mapping theorem, we have 
$$
T_n\overset{\mathcal{D}}{\rightarrow} \sup_{k=4,\cdots,n-4}\frac{nG_n^2(\frac{k}{n};\frac{1}{n},1)}{\sum_{t=2}^{k-1}G_n^2(\frac{t}{n};\frac{1}{n},\frac{k}{n})+\sum_{t=k+2}^{n-2}G_n^2(\frac{t}{n};\frac{k+1}{n},1)}.
$$

(ii) The proof is a simplified version of the proof of Theorem \ref{thm_rsrm} (ii), hence omitted here.

\qed
\vspace{4mm}

\subsection{Proof of Theorem \ref{thm_fix}}
Clearly, 
$$
T_n^{(s)}=\sup_{k=4,\cdots,n-4}\frac{(D^{(s)}(k;1,n))^2}{W_n^{(s)}(k;1,n)}\geq \frac{(D^{(s)}(k^*;1,n))^2}{W_n^{(s)}(k^*;1,n)}, 
$$
and 
$$
T_n=\sup_{k=4,\cdots,n-4}\frac{(D(k;1,n))^2}{W_n(k;1,n)}\geq \frac{(D(k^*;1,n))^2}{W_n(k^*;1,n)}, 
$$

Note that 
$W_n^{(s)}(k;1,n)=\frac{1}{n}\sum_{t=2}^{k^*-2}D^{(s)}(t;1,k^*)^2+\frac{1}{n}\sum_{t=k^*+2}^{n-2}D^{(s)}(t;k^*+1,n)^2.$ The construction of $D^{(s)}(t;1,k^*)^2$ (or  $D^{(s)}(t;k^*+1,n)^2$) only uses sample before (or after) the change point, so the change point has no influence on this part. The proof of Theorem \ref{thm_main} indicates that $4p^2n^{-6}\|\Sigma\|_F^{-2}W_n^{(s)}(k;1,n)=O_p(1)$ and similarly $4n^{-6}\|\Sigma\|_F^{-2}W_n(k;1,n)=O_p(1)$. Hence, it suffices to show $pn^{-3}\|\Sigma\|_F^{-1}D^{(s)}(k^*;1,n)\to_p\infty$ and $n^{-3}\|\Sigma\|_F^{-1}D(k^*;1,n)\to_p\infty$. 

Denote $\delta_i$ as the $i$th element of $\delta$.  By (\ref{decomp}), for $1\leq j_1\neq j_3\leq k^*$ and $k^*+1\leq j_2\neq j_4\leq n$, 
\begin{flalign*}
	p^{-1}\|Y_{j_1}-Y_{j_2}\|^2=& p^{-1}\|\delta\|^2+p^{-1}\sum_{i=1}^{p}(X_{j_1,i}-X_{j_2,i})^2-p^{-1}\sum_{i=1}^{p}2\delta_i(X_{j_1,i}-X_{j_2,i}),\\
	p^{-1}\|Y_{j_3}-Y_{j_4}\|^2=& p^{-1}\|\delta\|^2+p^{-1}\sum_{i=1}^{p}(X_{j_3,i}-X_{j_4,i})^2-p^{-1}\sum_{i=1}^{p}2\delta_i(X_{j_3,i}-X_{j_4,i}),\\
\end{flalign*}
and
\begin{flalign*}
	p^{-1}(Y_{j_1}-Y_{j_2})'(Y_{j_3}-Y_{j_4})=& p^{-1}\|\delta\|^2+p^{-1}\sum_{i=1}^{p}(X_{j_1,i}-X_{j_2,i})(X_{j_3,i}-X_{j_4,i})\\&-p^{-1}\sum_{i=1}^{p}\delta_i(X_{j_1,i}-X_{j_2,i})-p^{-1}\sum_{i=1}^{p}\delta_i(X_{j_3,i}-X_{j_4,i}).
\end{flalign*}

Using Theorem 8.2.2 in \cite{zhengyan1997limit}, and the independence of $X_i$'s, we have
\begin{flalign*}
	p^{-1}\|Y_{j_1}-Y_{j_2}\|^2\to_p\iota^2+2\sigma^2,\\
	p^{-1}\|Y_{j_3}-Y_{j_4}\|^2\to_p\iota^2+2\sigma^2,\\
\end{flalign*}
and
\begin{flalign*}
	p^{-1}(Y_{j_1}-Y_{j_2})'(Y_{j_3}-Y_{j_4})\to_p \iota^2.
\end{flalign*}

If $\iota>0$, then
$$
n^{-4}D^{(s)}(k^*;1,n)\to_p n^{-4}k^*(k^*-1)(n-k^*)(n-k^*-1) \frac{\iota^2}{\iota^2+2\sigma^2}>0,
$$
and 
$$
p^{-1}n^{-4}D(k^*;1,n)\to_p n^{-4}(k^*)(k^*-1)(n-k^*)(n-k^*-1)\iota^2>0.
$$
Hence, $$pn^{-3}\|\Sigma\|_F^{-1}D^{(s)}(k^*;1,n)=(pn\|\Sigma\|_F^{-1})n^{-4}D^{(s)}(k^*;1,n)\to_p\infty,$$ and $$n^{-3}\|\Sigma\|_F^{-1}D(k^*;1,n)=(pn\|\Sigma\|_F^{-1})p^{-1}n^{-4}D(k^*;1,n)\to_p\infty.$$

\qed
\vspace{4mm}

\subsection{Proof of Theorem \ref{thm_power}}
By symmetry, we only consider the case $l<k\leq k^*<m$. Since  under Assumption \ref{ass_power}, (\ref{consis}) still holds by Cauchy-Schwartz inequality, then using similar arguments in the proof of Theorem \ref{thm_main}, we have  
\begin{flalign}\label{D1_decomp}
	\begin{split}
		&2p\sigma^2D^{(s)}(k;l,m)\\=&\sum_{\substack{l\leq j_1,j_3\leq k\\j_1\neq j_3}}\sum_{\substack{k+1\leq j_2,j_4\leq m\\j_2\neq j_4}}(X_{j_1}-X_{j_2})'(X_{j_3}-X_{j_4})(1+o(1))
		\\&+(k-l+1)(k-l)(m-k^*)(m-k^*-1)\|\delta\|^2(1+o(1))\\&-\Big(2(k-l)(m-k^*)(m-k-2)\sum_{j=l}^{k}X_j'\delta+4(k-l)(k-l-1)(m-k^*)\sum_{j=k+1}^{k^*}X_j'\delta\Big)(1+o(1))
		\\:=&D^{(s)}_{(1)}(k;l,m)+D^{(s)}_{(2)}(k;l,m)-D^{(s)}_{(3)}(k;l,m).
	\end{split}
\end{flalign}
That is,  $2p\sigma^2D^{(s)}(k;l,m)=D(k;l,m)(1+o(1))$ for any triplet $(k,l,m)$, hence it suffices to consider $T_n^{(s)}$ as the results of $T_n$ are similar.

We first note that 
$$
\mathrm{Var}(X_i'\delta)=\delta'\Sigma\delta=o(\|\Sigma\|_F^2),
$$
hence by Chebyshev inequality, for any triplet $(k,l,m)$, we have 
\begin{equation}\label{D3}
	n^{-3}\|\Sigma\|_F^{-1}D^{(s)}_{(3)}(k;l,m)=o_p(1).
\end{equation}

(i)
By similar arguments in the proof of Theorem \ref{thm_fix}, it   suffices to show $$2p\sigma^2n^{-3}\|\Sigma\|_F^{-1}D^{(s)}(k^*;1,n)\to_p\infty.$$ 
In fact, by similar arguments used in the proof of Theorem \ref{thm_main}, we can show that $$	n^{-3}\|\Sigma\|_F^{-1}D^{(s)}_{(1)}(k;l,m)=O_p(1).$$ Then, recall (\ref{D1_decomp}), the result follows by noting  $$n^{-3}\|\Sigma\|_{F}^{-1} D^{(s)}_{(2)}\left(k^{*} ; 1, n\right)=n^{-3}\|\Sigma\|_{F}^{-1}(k-l+1)(k-l)\left(m-k^{*}\right)\left(m-k^{*}-1\right)\|\delta\|^{2}(1+o(1)) \rightarrow \infty.$$


(ii) As $n\|\Sigma\|_F^{-1}\|\delta\|^2\to0$, it follows from the same argument as (\ref{equiv}).

(iii) As $n\|\Sigma\|_F^{-1}\|\delta\|^2\to c_n\in (0,\infty)$, then we have 
\begin{flalign*}
	&n^{-3}\|\Sigma\|_F^{-1}D^{(s)}_{(2)}(k^*;m,l)]\\=&n^{-3}\|\Sigma\|_F^{-1}(k-l+1)(k-l)(m-k^*)(m-k^*-1)\|\delta\|^2(1+o(1))\\\to& c_n\frac{4{k-l+1\choose 2}{m-k^*\choose 2}}{n^4}.
\end{flalign*}

Therefore, continuous mapping theorem together with Lemma \ref{lem_fix} indicate that 
$$
T_n^{(s)}\overset{\mathcal{D}}{\rightarrow} \sup_{k=4,\cdots,n-4}\frac{n[\sqrt{2}G_n(\frac{k}{n};\frac{1}{n},1)+c_n\Delta_n(\frac{k}{n};\frac{1}{n},1)]^2}{\sum_{t=2}^{k-2}[\sqrt{2}G_n(\frac{t}{n};\frac{1}{n},\frac{k}{n})+c_n\Delta_n(\frac{t}{n};\frac{1}{n},\frac{k}{n})]^2+\sum_{t=k+2}^{n-2}[\sqrt{2}G_n(\frac{t}{n};\frac{k+1}{n},1)+c_n\Delta_n(\frac{t}{n};\frac{k+1}{n},1)]^2}.
$$

The last part of the proof is similar to the proof of Theorem \ref{thm_rsrmpower} (ii) below, and is  simpler, hence omitted.

\qed
\vspace{4mm}

\subsection{Proof of Theorem \ref{thm_rsrm}}
(i)
Note that 
\begin{equation}
	\frac{1}{p}\|Y_i-Y_j\|^2=\frac{1}{p}\sum_{\ell=1}^{p}(\frac{X_{i,\ell}}{R_i}-\frac{X_{j,\ell}}{R_j})^2,
\end{equation}
hence given $\mathcal{R}_n$, as $p\to\infty$, we have almost surely
\begin{equation}\label{consis2}
	\frac{1}{p}\|Y_i-Y_j\|^2\to \sigma^2(R_{i}^{-2}+R_{j}^{-2}).
\end{equation}
Note that
\begin{flalign}\label{D34}
	\begin{split}
		&p\sigma^2D^{(s)}(k;l,m)\\=&\sum_{\substack{l\leq j_1,j_3\leq k\\j_1\neq j_3}}\sum_{\substack{k+1\leq j_2,j_4\leq m\\j_2\neq j_4}}\frac{(Y_{j_1}-Y_{j_2})'(Y_{j_3}-Y_{j_4})}{(R_{j_1}^{-2}+R_{j_2}^{-2})^{1/2}(R_{j_3}^{-2}+R_{j_4}^{-2})^{1/2}}
		\\&+\sum_{\substack{l\leq j_1,j_3\leq k\\j_1\neq j_3}}\sum_{\substack{k+1\leq j_2,j_4\leq m\\j_2\neq j_4}}\frac{(Y_{j_1}-Y_{j_2})'(Y_{j_3}-Y_{j_4})}{(R_{j_1}^{-2}+R_{j_2}^{-2})^{1/2}(R_{j_3}^{-2}+R_{j_4}^{-2})^{1/2}}\Big\{\frac{p\sigma^2(R_{j_1}^{-2}+R_{j_2}^{-2})^{1/2}(R_{j_3}^{-2}+R_{j_4}^{-2})^{1/2}}{\|Y_{j_1}-Y_{j_2}\|\|Y_{j_3}-Y_{j_4}\|}-1\Big\}
		\\=:&[D_3(k;l,m)+D_4(k;l,m)].
	\end{split}
\end{flalign}
{Let 
	\begin{flalign*}
		A_{j_1,j_3}(k;l,m)=&\sum_{\substack{k+1\leq j_2,j_4\leq m\\j_2\neq j_4}}(R_{j_1}^{-2}+R_{j_2}^{-2})^{-1/2}(R_{j_3}^{-2}+R_{j_4}^{-2})^{-1/2},\\
		B_{j_2,j_4}(k;l,m)=&\sum_{\substack{l\leq j_1,j_3\leq k\\j_1\neq j_3}}(R_{j_1}^{-2}+R_{j_2}^{-2})^{-1/2}(R_{j_3}^{-2}+R_{j_4}^{-2})^{-1/2},\\
	\end{flalign*}
	and
	\begin{flalign*}
		C_{j_1,j_2}(k;l,m)=&-2\sum_{\substack{l\leq j_3\leq k\\j_3\neq j_1}}\sum_{\substack{k+1\leq j_4\leq m\\j_4\neq j_2}}(R_{j_1}^{-2}+R_{j_2}^{-2})^{-1/2}(R_{j_3}^{-2}+R_{j_4}^{-2})^{-1/2}.
	\end{flalign*}
	Then under $H_0$,
	\begin{align*}
		&D_3(k;l,m)\\
		=& \sum_{\substack{l \leq j_1, j_3 \leq k\\j_1 \neq j_3}}X_{j_1}^TX_{j_3}(R_{j_1}R_{j_3})^{-1}A_{j_1,j_3}(k;l,m)+\sum_{\substack{k + 1 \leq j_2, j_4 \leq m\\ j_2 \neq j_4}}X_{j_2}^TX_{j_4}(R_{j_2}R_{j_4})^{-1}B_{j_2,j_4}(k;l,m)\\
		&+\sum_{l \leq j_1 \leq k}\sum_{k+1 \leq j_2 \leq m}X_{j_1}^TX_{j_2}(R_{j_1}R_{j_2})^{-1}C_{j_1,j_2}(k;l,m).
	\end{align*}
	Denote that $U_1 = (X_1^TX_2,...,X_1^TX_n, X_2^TX_3,...,X_2^TX_n,...,X_{n-1}^TX_n)^T$ which contains all inner products of $X_i$ and $X_j$ for all $i\neq j$, and $U_2 = (R_1,...,R_n)^T$. By definition, $\sigma(U_1) \perp \!\!\! \perp \sigma(U_2)$, where $\sigma(U)$ is the $\sigma-$field generated by $U$, and we further observe that $2p\sigma^2D_3(k;l,m)$ is a continuous functional of $U_1$ and $U_2$. Hence to derive the limiting distribution of $2p\sigma^2D_3(k;l,m)$ when $p \rightarrow \infty$, it suffices to derive the limiting distribution of $(U_1,U_2)^T$.

	For any $\alpha \in \mathbb{R}^{n(n-1)/2}$, similar to the proof of Theorem \ref{thm_main}, by Theorem 4.0.1 in \cite{zhengyan1997limit} we have
	$$\|\Sigma\|_F^{-1}\alpha^TU_1 \overset{\mathcal{D}}{\rightarrow} \alpha^T\mathcal{Z} := \alpha^T(\mathcal{Z}_{1,2},\mathcal{Z}_{1,3},...,\mathcal{Z}_{1,n},\mathcal{Z}_{2,3},...,\mathcal{Z}_{2,n},...,\mathcal{Z}_{n-1,n})^T,$$
	where $\mathcal{Z}_{1,2},...,\mathcal{Z}_{n-1,n}$ are i.i.d. standard normal random variables, and we can assume $\mathcal{Z}$ is independent of $U_2$. For the ease of our notation, we let $\mathcal{Z}_{i,j} = \mathcal{Z}_{j,i}$, for all $i > j$. Furthermore since $\sigma(U_1) \perp \!\!\! \perp \sigma(U_2)$, for any $\alpha \in \mathbb{R}^{n(n-1)/2}$ and $\beta \in \mathbb{R}^{n}$, the characteristic function of $\alpha^TU_1 + \beta^TU_2$ is the product of the characteristic function of $\alpha^TU_1$ and that of $\beta^TU_2$. By applying the Cram\'er-Wold device, $(\|\Sigma\|_F^{-1}U_1,U_2) \overset{\mathcal{D}}{\rightarrow} (\mathcal{Z},U_2)$.
	Therefore, by continuous mapping theorem,  as $p \rightarrow \infty$, 
	\begin{equation}\label{D3_con}
		n^{-3}\|\Sigma\|_F^{-1}D_3(k;l,m) \overset{\mathcal{D}}{\rightarrow} G_n^{(\mathcal{R}_n,s)}(k/n;l/n,m/n),
	\end{equation}
	where
	\begin{flalign}\label{GnR}
		\begin{split}
			&G_n^{(\mathcal{R}_n,s)}(k/n;l/n,m/n) \\:=& n^{-3}\sum_{\substack{l \leq j_1, j_3 \leq k\\j_1 \neq j_3}}\mathcal{Z}_{j_1,j_3}(R_{j_1}R_{j_3})^{-1}A_{j_1,j_3}(k,l,m)+n^{-3}\sum_{\substack{k + 1 \leq j_2, j_4 \leq m\\ j_2 \neq j_4}}\mathcal{Z}_{j_2,j_4}(R_{j_2}R_{j_4})^{-1}B_{j_2,j_4}(k,l,m)\\
			& +n^{-3}\sum_{l \leq j_1 \leq k}\sum_{k+1 \leq j_2 \leq m}\mathcal{Z}_{j_1,j_2}(R_{j_1}R_{j_2})^{-1}C_{j_1,j_2}(k,l,m).
		\end{split}
	\end{flalign}
	
	It is clear that the conditional distribution of $G_n^{(\mathcal{R}_n,s)}(k/n;l/n,m/n)$ given $\mathcal{R}_n$ is Gaussian, and   for any $l_1 < k_1 < m_1, l_2 < k_2 < m_2$, $k_1,k_2,l_1,l_2,m_1,m_2 = 1,2,...,n$, the covariance structure is given by
	\begin{flalign}\label{cov}
		&\mathrm{Cov}(G_n^{(\mathcal{R}_n,s)}(k_1/n;l_1/n,m_1/n),G_n^{(\mathcal{R}_n,s)}(k_2/n;l_2/n,m_2/n)|\mathcal{R}_n) 
		\\=&\notag2n^{-6}\Big\{\sum_{\substack{(l_1\lor l_2)\leq j_1,j_2\leq (k_1\land k_2)\\j_1\neq j_2}}R_{j_1}^{-2}R_{j_2}^{-2}A_{j_1,j_2}(k_1;l_1,m_1)A_{j_1,j_2}(k_2;l_2,m_2)
		\\&\notag+\sum_{\substack{(l_1\lor k_2+1)\leq j_1,j_2\leq (k_1\land m_2)\\j_1\neq j_2}}R_{j_1}^{-2}R_{j_2}^{-2}A_{j_1,j_2}(k_1;l_1,m_1)B_{j_1,j_2}(k_2;l_2,m_2)
		\\&\notag+2\sum_{j_1=(l_1\lor l_2)}^{k_2}\sum_{j_2=k_2+1}^{(m_2\land k_1)}\mathbf{1}(k_1>k_2)R_{j_1}^{-2}R_{j_2}^{-2}A_{j_1,j_2}(k_1;l_1,m_1)C_{j_1,j_2}(k_2;l_2,m_2)
		\\&\notag+\sum_{\substack{(k_1+1\lor l_2)\leq j_1,j_2\leq (m_1\land k_2)\\j_1\neq j_2}}R_{j_1}^{-2}R_{j_2}^{-2}B_{j_1,j_2}(k_1;l_1,m_1)A_{j_1,j_2}(k_2;l_2,m_2)
		\\&\notag+\sum_{\substack{(k_1+1\lor k_2+1)\leq j_1,j_2\leq (m_1\land m_2)\\j_1\neq j_2}}R_{j_1}^{-2}R_{j_2}^{-2}B_{j_1,j_2}(k_1;l_1,m_1)B_{j_1,j_24}(k_2;l_2,m_2)
		\\&\notag+2\sum_{j_1=(k_1+1\lor l_2)}^{k_2}\sum_{j_2=k_2+1}^{(m_1\land m_2)}\mathbf{1}(m_1> k_2)R_{j_1}^{-2}R_{j_2}^{-2}B_{j_1,j_2}(k_1;l_1,m_1)C_{j_1,j_2}(k_2;l_2,m_2)
		\\&\notag+2\sum_{j_1=(l_1\lor l_2)}^{k_1}\sum_{j_2=k_1+1}^{(m_1\land k_2)}\mathbf{1}(k_2>k_1)R_{j_1}^{-2}R_{j_2}^{-2}C_{j_1,j_2}(k_1;l_1,m_1)A_{j_1,j_2}(k_2;l_2,m_2)
		\\&\notag+2\sum_{j_1=(k_2+1\lor l_1)}^{k_1}\sum_{j_2=k_1+1}^{(m_1\land m_2)}\mathbf{1}(m_2> k_1)R_{j_1}^{-2}R_{j_2}^{-2}C_{j_1,j_2}(k_1;l_1,m_1)B_{j_1,j_2}(k_2;l_2,m_2)
		\\&\notag+\sum_{j_1=(l_1\lor l_2)}^{(k_1\land k_2)}\sum_{j_2=(k_1+1\lor k_2+1)}^{(m_1\land m_2)}R_{j_1}^{-2}R_{j_2}^{-2}C_{j_1,j_2}(k_1;l_1,m_1)C_{j_1,j_2}(k_2;l_2,m_2)\Big\}.
	\end{flalign}
	Clearly, when $R_i\equiv1$, we have $2D_3(k;l,m)=D_1(k;l,m)$ where $D_1(k;l,m)$ is defined in (\ref{D}), and the result reduces to (\ref{D1}). 
	
	Using (\ref{consis2}), we can see that given $\mathcal{R}_n$, $\frac{D_4(k;l,m)}{n^3\|\Sigma\|_F}=o_p(1)$. Hence, given $\mathcal{R}_n$, we have 
	$$
	T_n^{(s)}=\sup_{k=4,\cdots,n-4}\frac{[D_3(k;1,n)]^2}{\frac{1}{n}\sum_{t=l+1}^{k-2}D_3(t;l,k)^2+\frac{1}{n}\sum_{t=k+2}^{m-2}D_3(t;k+1,m)^2}+o_p(1).
	$$
	Then, by (\ref{D3_con}), we have that as $p\to\infty$,
	$$
	T_n^{(s)}|\mathcal{R}_n\overset{\mathcal{D}}{\rightarrow}\mathcal{T}_n^{(\mathcal{R}_n,s)}:= \sup_{k=4,\cdots,n-4}\frac{n[G_n^{(\mathcal{R}_n,s)}(k/n;1/n,1)]^2}{\sum_{t=2}^{k-1}[G_n^{(\mathcal{R}_n,s)}(t/n;1/n,k/n)]^2+\sum_{t=k+2}^{n-2}[G_n^{(\mathcal{R}_n,s)}(t/n;(k+1)/n,1)]^2}.$$ 
	
	As for $T_n$, note that \begin{align*}
		D(k;l,m)
		=& (m-k)(m-k-1)\sum_{\substack{l \leq j_1, j_3 \leq k\\j_1 \neq j_3}}X_{j_1}^TX_{j_3}(R_{j_1}R_{j_3})^{-1}\\&+(k-l+1)(k-l)\sum_{\substack{k + 1 \leq j_2, j_4 \leq m\\ j_2 \neq j_4}}X_{j_2}^TX_{j_4}(R_{j_2}R_{j_4})^{-1}\\
		&-2(k-l)(m-k-1)\sum_{l \leq j_1 \leq k}\sum_{k+1 \leq j_2 \leq m}X_{j_1}^TX_{j_2}(R_{j_1}R_{j_2})^{-1}.
	\end{align*}
	Using similar arguments as in (\ref{D3_con}), we have 
	\begin{equation}\label{D_con}
		n^{-3}\|\Sigma\|_F^{-1}D(k;l,m) \overset{\mathcal{D}}{\rightarrow} G_n^{(\mathcal{R}_n)}(k/n;l/n,m/n),
	\end{equation}
	where
	\begin{flalign}\label{GnR2}
		\begin{split}
			G_n^{(\mathcal{R}_n)}(k/n;l/n,m/n)=& (m-k)(m-k-1)n^{-3}\sum_{\substack{l \leq j_1, j_3 \leq k\\j_1 \neq j_3}}\mathcal{Z}_{j_1,j_3}(R_{j_1}R_{j_3})^{-1}\\&+(k-l+1)(k-l)n^{-3}\sum_{\substack{k + 1 \leq j_2, j_4 \leq m\\ j_2 \neq j_4}}\mathcal{Z}_{j_2,j_4}(R_{j_2}R_{j_4})^{-1}\\
			& -2(k-l)(m-k-1)n^{-3}\sum_{l \leq j_1 \leq k}\sum_{k+1 \leq j_2 \leq m}\mathcal{Z}_{j_1,j_2}(R_{j_1}R_{j_2})^{-1}.
		\end{split}
	\end{flalign}
	Similar to  $G_n^{(\mathcal{R}_n,s)}(k/n;l/n,m/n)$, the conditional distribution of $G_n^{(\mathcal{R}_n)}(k/n;l/n,m/n)$ given $\mathcal{R}_n$ is Gaussian, and   for any $l_1 < k_1 < m_1, l_2 < k_2 < m_2$, $k_1,k_2,l_1,l_2,m_1,m_2 = 1,2,...,n$, the covariance structure is given by
	\begin{flalign}\label{cov2}
		&\mathrm{Cov}(G_n^{(\mathcal{R}_n)}(k_1/n;l_1/n,m_1/n),G_n^{(\mathcal{R}_n)}(k_2/n;l_2/n,m_2/n)|\mathcal{R}_n) 
		\\=&\notag8n^{-6}\Big\{\sum_{\substack{(l_1\lor l_2)\leq j_1,j_2\leq (k_1\land k_2)\\j_1\neq j_2}}R_{j_1}^{-2}R_{j_2}^{-2}{m_1-k_1\choose 2}{m_2-k_2\choose 2}
		\\&\notag+\sum_{\substack{(l_1\lor k_2+1)\leq j_1,j_2\leq (k_1\land m_2)\\j_1\neq j_2}}R_{j_1}^{-2}R_{j_2}^{-2}{m_1-k_1\choose 2}{k_1-l_1+1\choose 2}
		\\&\notag-2\sum_{j_1=(l_1\lor l_2)}^{k_2}\sum_{j_2=k_2+1}^{(m_2\land k_1)}\mathbf{1}(k_1>k_2)R_{j_1}^{-2}R_{j_2}^{-2}{m_1-k_1\choose 2}(k_2-l_2)(m_2-k_2-1)
		\\&\notag+\sum_{\substack{(k_1+1\lor l_2)\leq j_1,j_2\leq (m_1\land k_2)\\j_1\neq j_2}}R_{j_1}^{-2}R_{j_2}^{-2}{k_1-l_1+1\choose2}{m_2-k_2\choose 2}
		\\&\notag+\sum_{\substack{(k_1+1\lor k_2+1)\leq j_1,j_2\leq (m_1\land m_2)\\j_1\neq j_2}}R_{j_1}^{-2}R_{j_2}^{-2}{k_1-l_1+1\choose2}{k_2-l_2+1\choose2}
		\\&\notag-2\sum_{j_1=(k_1+1\lor l_2)}^{k_2}\sum_{j_2=k_2+1}^{(m_1\land m_2)}\mathbf{1}(m_1> k_2)R_{j_1}^{-2}R_{j_2}^{-2}B_{j_1,j_2}{k_1-l_1+1\choose2}(k_2-l_2)(m_2-k_2-1)
		\\&\notag-2\sum_{j_1=(l_1\lor l_2)}^{k_1}\sum_{j_2=k_1+1}^{(m_1\land k_2)}\mathbf{1}(k_2>k_1)R_{j_1}^{-2}R_{j_2}^{-2}(k_1-l_1)(m_1-k_1-1){m_2-k_2\choose 2}
		\\&\notag-2\sum_{j_1=(k_2+1\lor l_1)}^{k_1}\sum_{j_2=k_1+1}^{(m_1\land m_2)}\mathbf{1}(m_2> k_1)R_{j_1}^{-2}R_{j_2}^{-2}(k_1-l_1)(m_1-k_1-1){k_2-l_2+1\choose2}
		\\&\notag+\sum_{j_1=(l_1\lor l_2)}^{(k_1\land k_2)}\sum_{j_2=(k_1+1\lor k_2+1)}^{(m_1\land m_2)}R_{j_1}^{-2}R_{j_2}^{-2}(k_1-l_1)(m_1-k_1-1)(k_2-l_2)(m_2-k_2-1)\Big\}.
	\end{flalign}
	Hence, as $p\to\infty$,
	$$
	T_n|\mathcal{R}_n\overset{\mathcal{D}}{\rightarrow}\mathcal{T}_n^{(\mathcal{R}_n)}:= \sup_{k=4,\cdots,n-4}\frac{n[G_n^{(\mathcal{R}_n)}(k/n;1/n,1)]^2}{\sum_{t=2}^{k-1}[G_n^{(\mathcal{R}_n)}(t/n;1/n,k/n)]^2+\sum_{t=k+2}^{n-2}[G_n^{(\mathcal{R}_n)}(t/n;(k+1)/n,1)]^2}.$$ 
	
	(ii)

	{We shall only show the process convergence $G_n^{(\mathcal{R}_n,s)}(\cdot) \rightsquigarrow \E\Big[\frac{R_1R_2}{\sqrt{(R_1^2+R_3^2)(R_2^2+R_3^2)}}\Big]\sqrt{2}G(\cdot)$. That $G_n^{(\mathcal{R}_n)}(\cdot) \rightsquigarrow \E(R_1^{-2})\sqrt{2}G(\cdot)$ is similar and simpler. 
		Once the process convergence is obtained, the limiting distributions of $\mathcal{T}_n^{(\mathcal{R}_n,s)}$ and $\mathcal{T}_n^{(\mathcal{R}_n)}$ can be easily obtained by the continuous mapping theorem.   
		
		The proof for the process convergence contains two parts: the finite dimensional convergence and the tightness.
		
		To show the finite dimensional convergence, we need to show that for any positive integer $N$, any fixed $u_1,u_2,...,u_N \in [0,1]^3$ and any $\alpha_1,...,\alpha_N \in \mathbb{R}$,
		$$\alpha_1G_n^{(\mathcal{R}_n,s)}(u_1) + \cdots \alpha_NG_n^{(\mathcal{R}_n,s)}(u_N)\overset{\mathcal{D}}{\rightarrow}\E^2\Big[\frac{R_1R_2}{\sqrt{(R_1^2+R_3^2)(R_2^2+R_3^2)}}\Big]\sqrt{2}[\alpha_1G(u_1) + \cdots \alpha_NG(u_N)],$$
		where  for $u = (u^{(1)},u^{(2)},u^{(3)})^T$, $G_n(u) = G_n(u_1;u_2,u_3)$.  Since both $G_n^{(\mathcal{R}_n,s)}(\cdot)|\mathcal{R}_n$ and $G(\cdot)$ are Gaussian processes, by Lemma \ref{lem_equiv} we have 
		\begin{flalign*}
			&P(\alpha_1G_n^{(\mathcal{R}_n,s)}(u_1) + \cdots \alpha_kG_n^{(\mathcal{R}_n,s)}(u_k) < x|\mathcal{R}_n)\\\to_p
			&P(\E\Big[\frac{R_1R_2}{\sqrt{(R_1^2+R_3^2)(R_2^2+R_3^2)}}\Big]\sqrt{2}[\alpha_1G(u_1) + \cdots \alpha_NG(u_N)] < x).
		\end{flalign*}
		Then by bounded convergence theorem we have
		\begin{align*}
			&\lim_{n\rightarrow\infty}P(\alpha_1G_n^{(\mathcal{R}_n,s)}(u_1) + \cdots \alpha_kG_n^{(\mathcal{R}_n,s)}(u_k) < x)\\ =& \lim_{n\rightarrow\infty}\E[P(\alpha_1G_n^{(\mathcal{R}_n,s)}(u_1) + \cdots \alpha_kG_n^{(\mathcal{R}_n,s)}(u_k) < x|\mathcal{R}_n)]\\
			=&\E[\lim_{n\rightarrow\infty}P(\alpha_1G_n^{(\mathcal{R}_n,s)}(u_1) + \cdots \alpha_kG_n^{(\mathcal{R}_n,s)}(u_k) < x|\mathcal{R}_n)]\\ =&\E \Big[\frac{R_1R_2}{\sqrt{(R_1^2+R_3^2)(R_2^2+R_3^2)}}\Big]\sqrt{2}[\alpha_1G(u_1) + \cdots \alpha_kG(u_k)] < x)]\\
			=&P(\E\Big[\frac{R_1R_2}{\sqrt{(R_1^2+R_3^2)(R_2^2+R_3^2)}}\Big]\sqrt{2}[\alpha_1G(u_1) + \cdots \alpha_kG(u_k)] < x).
		\end{align*}
		This completes the proof of the finite dimensional convergence. 
		
		To show the tightness, it suffices to show that there exists $C > 0$ such that $$\E[(G_n^{(\mathcal{R}_n,s)}(u) - G_n^{(\mathcal{R}_n,s)}(v))^8] \leq C(\|u - v\|^4 + 1/n^4),$$ for any $u,v \in [0,1]^3$ (see the proof of equation S8.12 in \cite{wang2021inference}).
		
		Since given $\mathcal{R}_n$, $G_n^{(\mathcal{R}_n,s)}(\cdot)$ is a Gaussian process, we have
		\begin{align*}
			&\E[(G_n^{(\mathcal{R}_n,s)}(u) - G_n^{(\mathcal{R}_n,s)}(v))^8] = \E[\E[(G_n^{(\mathcal{R}_n,s)}(u) - G_n^{(\mathcal{R}_n,s)}(v))^8|\mathcal{R}_n]]\\
			&= C\E[\mathrm{Var}((G_n^{(\mathcal{R}_n,s)}(u) - G_n^{(\mathcal{R}_n,s)}(v))|\mathcal{R}_n)^4].
		\end{align*}
		
		By (\ref{cov}), for $u = (k_1/n,l_1/n,m_1/n)$ (and similar for $v = (k_2/n, l_2/n, m_2/n)$) this reduces to
		\begin{align*}
			&\mathrm{Var}(G_n^{(\mathcal{R}_n,s)}(u)|\mathcal{R}_n)\\
			=& 2n^{-6}\Big\{\sum_{\substack{l_1 \leq j_1, j_3 \leq k_1\\j_1 \neq j_3}}(R_{j_1}R_{j_3})^{-2}A_{j_1,j_3}(k_1,l_1,m_1)^2 +\sum_{\substack{k_1 + 1 \leq j_2, j_4 \leq m_1\\ j_2 \neq j_4}}(R_{j_2}R_{j_4})^{-2}B_{j_2,j_4}(k_1,l_1,m_1)^2\\
			&+\sum_{l_1 \leq j_1 \leq k_1}\sum_{k_1+1 \leq j_2 \leq m_1}(R_{j_1}R_{j_2})^{-2}C_{j_1,j_2}(k_1,l_1,m_1)^2\Big\}.
		\end{align*}
		
		Note that 
		\begin{align*}
			&\E[(G_n^{(\mathcal{R}_n,s)}(k_1/n;l_1/n,m_1/n) - G_n^{(\mathcal{R}_n,s)}(k_2/n;l_2/n,m_2/n))^8] \\
			\lesssim& \E[(G_n^{(\mathcal{R}_n,s)}(k_1/n;l_1/n,m_1/n) - G_n^{(\mathcal{R}_n,s)}(k_2/n;l_1/n,m_1/n))^8] \\
			&+ \E[(G_n^{(\mathcal{R}_n,s)}(k_2/n;l_1/n,m_1/n) - G_n^{(\mathcal{R}_n,s)}(k_2/n;l_2/n,m_1/n))^8]\\
			&+ \E[(G_n^{(\mathcal{R}_n,s)}(k_2/n;l_2/n,m_1/n) - G_n^{(\mathcal{R}_n,s)}(k_2/n;l_2/n,m_2/n))^8]\\
			=& I_1 + I_2 + I_3.
		\end{align*}
		
		We shall analyze $I_1$ first, and WLOG we let $k_1 < k_2$. Then we have (with $l_1=l_2,m_1=m_2$)
		\begin{flalign*}
			&\mathrm{Cov}(G_n^{(\mathcal{R}_n,s)}(k_1/n;l_1/n,m_1/n),G_n^{(\mathcal{R}_n,s)}(k_2/n;l_2/n,m_2/n)|\mathcal{R}_n)
			\\=&2n^{-6}\Big\{\sum_{\substack{l_1 \leq j_1,j_3\leq k_1 \\j_1\neq j_3}}R_{j_1}^{-2}R_{j_3}^{-2}A_{j_1,j_3}(k_1;l_1,m_1)A_{j_1,j_3}(k_2;l_1,m_1)
			\\&+\sum_{\substack{k_1+1\leq j_1,j_2\leq k_2\\j_1\neq j_2}}R_{j_1}^{-2}R_{j_2}^{-2}B_{j_1,j_2}(k_1;l_1,m_1)A_{j_1,j_2}(k_2;l_2,m_2)
			\\&+\sum_{\substack{k_2+1\leq j_2,j_4\leq m_1 \\j_2\neq j_4}}R_{j_2}^{-2}R_{j_4}^{-2}B_{j_2,j_4}(k_1;l_1,m_1)B_{j_2,j_4}(k_2;l_2,m_2)
			\\&+2\sum_{j_1=k_1+1}^{k_2}\sum_{j_2=k_2+1}^{m_1}R_{j_1}^{-2}R_{j_2}^{-2}C_{j_1,j_2}(k_2;l_2,m_2)B_{j_1,j_2}(k_1;l_1,m_1)
			\\&+2\sum_{j_1=l_1}^{k_1}\sum_{j_2=k_1+1}^{k_2}R_{j_1}^{-2}R_{j_2}^{-2}C_{j_1,j_2}(k_1;l_1,m_1)A_{j_1,j_2}(k_2;l_2,m_2)
			\\&+\sum_{j_1=l_1}^{k_1}\sum_{j_2=k_2+1}^{m_1}R_{j_1}^{-2}R_{j_2}^{-2}C_{j_1,j_2}(k_1;l_1,m_1)C_{j_1,j_2}(k_2;l_2,m_2)\Big\}.
		\end{flalign*}
		
		Hence, 
		\begin{align*}
			&\mathrm{Var}(G_n^{(\mathcal{R}_n,s)}(k_1/n;l_1/n,m_1/n)-G_n^{(\mathcal{R}_n,s)}(k_2/n;l_2/n,m_2/n)|\mathcal{R}_n)\\
			=& 2n^{-6}\Big\{\sum_{\substack{l_1 \leq j_1, j_3 \leq k_1\\j_1 \neq j_3}}(R_{j_1}R_{j_3})^{-2}A_{j_1,j_3}(k_1,l_1,m_1)^2 +\sum_{\substack{k_1 + 1 \leq j_2, j_4 \leq m_1\\ j_2 \neq j_4}}(R_{j_2}R_{j_4})^{-2}B_{j_2,j_4}(k_1,l_1,m_1)^2\\
			&+\sum_{l_1 \leq j_1 \leq k_1}\sum_{k_1+1 \leq j_2 \leq m_1}(R_{j_1}R_{j_2})^{-2}C_{j_1,j_2}(k_1,l_1,m_1)^2\Big\}\\
			+& 2n^{-6}\Big\{\sum_{\substack{l_1 \leq j_1, j_3 \leq k_2\\j_1 \neq j_3}}(R_{j_1}R_{j_3})^{-2}A_{j_1,j_3}(k_2,l_1,m_1)^2 +\sum_{\substack{k_2 + 1 \leq j_2, j_4 \leq m_1\\ j_2 \neq j_4}}(R_{j_2}R_{j_4})^{-2}B_{j_2,j_4}(k_2,l_1,m_1)^2\\
			&+\sum_{l_1 \leq j_1 \leq k_2}\sum_{k_2+1 \leq j_2 \leq m_1}(R_{j_1}R_{j_2})^{-2}C_{j_1,j_2}(k_2,l_1,m_1)^2\Big\}\\
			-&4n^{-6}\Big\{\sum_{\substack{l_1 \leq j_1,j_3\leq k_1 \\j_1\neq j_3}}R_{j_1}^{-2}R_{j_3}^{-2}A_{j_1,j_3}(k_1;l_1,m_1)A_{j_1,j_3}(k_2;l_1,m_1)
			\\&+\sum_{\substack{k_1+1\leq j_1,j_2\leq k_2\\j_1\neq j_2}}R_{j_1}^{-2}R_{j_2}^{-2}B_{j_1,j_2}(k_1;l_1,m_1)A_{j_1,j_2}(k_2;l_2,m_2)
			\\&+\sum_{\substack{k_2+1\leq j_2,j_4\leq m_1 \\j_2\neq j_4}}R_{j_2}^{-2}R_{j_4}^{-2}B_{j_2,j_4}(k_1;l_1,m_1)B_{j_2,j_4}(k_2;l_2,m_2)
			\\&+2\sum_{j_1=k_1+1}^{k_2}\sum_{j_2=k_2+1}^{m_1}R_{j_1}^{-2}R_{j_2}^{-2}C_{j_1,j_2}(k_2;l_2,m_2)B_{j_1,j_2}(k_1;l_1,m_1)
			\\&+2\sum_{j_1=l_1}^{k_1}\sum_{j_2=k_1+1}^{k_2}R_{j_1}^{-2}R_{j_2}^{-2}C_{j_1,j_2}(k_1;l_1,m_1)A_{j_1,j_2}(k_2;l_2,m_2)
			\\&+\sum_{j_1=l_1}^{k_1}\sum_{j_2=k_2+1}^{m_1}R_{j_1}^{-2}R_{j_2}^{-2}C_{j_1,j_2}(k_1;l_1,m_1)C_{j_1,j_2}(k_2;l_2,m_2)\Big\}.
		\end{align*}
		
		By rearranging terms we have
		\begin{align*}
			&\mathrm{Var}(G_n^{(\mathcal{R}_n,s)}(k_1/n;l_1/n,m_1/n)-G_n^{(\mathcal{R}_n,s)}(k_2/n;l_2/n,m_2/n)|\mathcal{R}_n)\\
			=& 2n^{-6}\Big\{\sum_{\substack{l_1 \leq j_1, j_3 \leq k_1\\j_1 \neq j_3}}(R_{j_1}R_{j_3})^{-2}(A_{j_1,j_3}(k_1,l_1,m_1) - A_{j_1,j_3}(k_2,l_1,m_1))^2\\
			&+\sum_{j_1 = k_1 + 1}^{k_2}\sum_{j_3 = l_1 }^{j_1-1}(R_{j_1}R_{j_3})^{-2}(A_{j_1,j_3}(k_2,l_1,m_1)^2 +A_{j_3,j_1}(k_2,l_1,m_1)^2)\\
			&+\sum_{\substack{k_2+1\leq j_2,j_4\leq m_1 \\j_2\neq j_4}}R_{j_2}^{-2}R_{j_4}^{-2}(B_{j_2,j_4}(k_1;l_1,m_1) - B_{j_2,j_4}(k_2;l_1,m_1))^2\\
			&+\sum_{j_2 = k_1 + 1}^{k_2-1}\sum_{j_4 = j_2 + 1}^{m_1}R_{j_2}^{-2}R_{j_4}^{-2}(B_{j_2,j_4}(k_1;l_1,m_1)^2 + B_{j_4,j_2}(k_1;l_1,m_1)^2)\\
			&+\sum_{j_1 = l_1}^{k_1}\sum_{j_2 = k_2 + 1}^{m_1}R_{j_1}^{-2}R_{j_2}^{-2}(C_{j_1,j_2}(k_1,l_1,m_1) - C_{j_1,j_2}(k_2,l_1,m_1))^2\\
			&+ \sum_{j_1 = l_1}^{k_1}\sum_{j_2 = k_1 + 1}^{k_2}R_{j_1}^{-2}R_{j_2}^{-2}C_{j_1,j_2}(k_1,l_1,m_1)^2 + \sum_{j_1 = k_1 + 1}^{k_2}\sum_{j_2 = k_2 + 1}^{m_1}R_{j_1}^{-2}R_{j_2}^{-2}C_{j_1,j_2}(k_2,l_1,m_1)^2\\
			&-2\sum_{\substack{k_1+1\leq j_1,j_2\leq k_2\\j_1\neq j_2}}R_{j_1}^{-2}R_{j_2}^{-2}B_{j_1,j_2}(k_1;l_1,m_1)A_{j_1,j_2}(k_2;l_1,m_1)\\
			&-4\sum_{j_1=k_1+1}^{k_2}\sum_{j_2=k_2+1}^{m_1}R_{j_1}^{-2}R_{j_2}^{-2}C_{j_1,j_2}(k_2;l_1,m_1)B_{j_1,j_2}(k_1;l_1,m_1)
			\\&-4\sum_{j_1=l_1}^{k_1}\sum_{j_2=k_1+1}^{k_2}R_{j_1}^{-2}R_{j_2}^{-2}C_{j_1,j_2}(k_1;l_1,m_1)A_{j_1,j_2}(k_2;l_1,m_1)\\
			=& \sum_{l = 1}^{10}J_i.
		\end{align*}
		
		Thus by CR-inequality we have $I_1 = \E[(\sum_{l = 1}^{10}J_l)^4] \lesssim \sum_{i = 1}^{10}\E[J_l^4]$. We shall analyze $\E[J_1^4]$ first. Note that 
		\begin{align*}
			&A_{j_1,j_3}(k_1,l_1,m_1) - A_{j_1,j_3}(k_2,l_1,m_1)\\
			= &\sum_{\substack{k_1+1\leq j_2,j_4\leq m_1\\j_2\neq j_4}}(R_{j_1}^{-2}+R_{j_2}^{-2})^{-1/2}(R_{j_3}^{-2}+R_{j_4}^{-2})^{-1/2}   -\sum_{\substack{k_2+1\leq j_2,j_4\leq m_1\\j_2\neq j_4}}(R_{j_1}^{-2}+R_{j_2}^{-2})^{-1/2}(R_{j_3}^{-2}+R_{j_4}^{-2})^{-1/2}\\
			=&\sum_{j_2 = k_1 + 1}^{k_2}\sum_{j_4 = j_2 + 1}^{m_1}(R_{j_1}^{-2}+R_{j_2}^{-2})^{-1/2}(R_{j_3}^{-2}+R_{j_4}^{-2})^{-1/2} + \sum_{j_4 = k_1 + 1}^{k_2}\sum_{j_2 = j_4 + 1}^{m_1}(R_{j_1}^{-2}+R_{j_2}^{-2})^{-1/2}(R_{j_3}^{-2}+R_{j_4}^{-2})^{-1/2}\\
			\leq& 2\sum_{j_2 = k_1 + 1}^{k_2}\sum_{j_4 = j_2 + 1}^{m_1}(2|R_{j_1}|^{-1}|R_{j_2}|^{-1})^{-1/2}(2|R_{j_3}|^{-1}|R_{j_4}|^{-1})^{-1/2}\\
			\leq&|R_{j_1}R_{j_3}|^{1/2}\sum_{j_2 = k_1 + 1}^{k_2}\sum_{j_4 = j_2 + 1}^{m_1}|R_{j_2}R_{j_4}|^{1/2}.
		\end{align*}
		Since $A_{j_1,j_3}(k_1,l_1,m_1) - A_{j_1,j_3}(k_2,l_1,m_1) > 0$ and $J_1 > 0$ almost surely, we have
		
			$$\E[J_1^4] \leq 2^4n^{-24}\E[\prod_{i = 1}^4(\sum_{\substack{l_1 \leq j_{1,i}, j_{3,i} \leq k_1\\j_{1,i} \neq j_{3,i}}}\sum_{j_{2,i} = k_1 + 1}^{k_2}\sum_{j_{4,i} = j_{2,i} + 1}^{m_1}\sum_{j_{2,i}' = k_1 + 1}^{k_2}\sum_{j_{4,i}' = j_{2,1} + 1}^{m_1}
			|R_{j_{1,i}}R_{j_{3,i}}|^{-1}|R_{j_{2,i}}R_{j_{4,i}}|^{1/2}|R_{j_{2,i}'}R_{j_{4,i}'}|^{1/2})].
			$$
		By the H\"older's inequalilty, and the fact that $j_{1,s} \neq j_{3,s}$, $j_{2,s} \neq j_{4,s}$, $j_{2,s}' \neq j_{4,s}'$ and $j_{1,s},j_{3,s}$ are not identical to any of $\{j_{2,s}, j_{4,s}, j_{2,s}, j_{4,s}\}$ for any $s = 1,2,3,4$, we have
		\begin{align*}
			&\E[|R_{j_{1,1}}R_{j_{3,1}}|^{-1}|R_{j_{2,1}}R_{j_{4,1}}|^{1/2}|R_{j_{2,1}'}R_{j_{4,1}'}|^{1/2}\cdots|R_{j_{1,4}}R_{j_{3,4}}|^{-1}|R_{j_{2,4}}R_{j_{4,4}}|^{1/2}|R_{j_{2,4}'}R_{j_{4,4}'}|^{1/2}]\\
			\leq&\prod_{s = 1}^4\E[((|R_{j_{1,s}}||R_{j_{3,s}}|)^{-1}|R_{j_{2,s}}R_{j_{4,s}}R_{j_{2,s}'}R_{j_{4,s}'}|^{1/2})^4]^{1/4}\\
			=&\prod_{s = 1}^4\E[(R_{j_{1,s}}R_{j_{3,s}})^{-4}R_{j_{2,s}}^{2}R_{j_{4,s}}^{2}R_{j_{2,s}'}^{2}R_{j_{4,s}'}^{2}]^{1/4} = \prod_{s = 1}^4\Big\{\E[R_{j_{1,s}}^{-4}]\E[R_{j_{3,s}}^{-4}]\E[R_{j_{2,s}}^{2}R_{j_{4,s}}^{2}R_{j_{2,s}'}^{2}R_{j_{4,s}'}^{2}]\Big\}^{1/4}\\
			\leq & \prod_{s = 1}^4\Big\{\E[R_{j_{1,s}}^{-4}]\E[R_{j_{3,s}}^{-4}]\E[R_{j_{2,s}}^{4}R_{j_{4,s}}^{4}]^{1/2}\E[R_{j_{2,s}'}^{4}R_{j_{4,s}'}^{4}]^{1/2}\Big\}^{1/4} = \E[R_{1}^{-4}]^2\E[R_{2}^{4}]^2.
		\end{align*}
		Therefore,
		\begin{align*}
			\E[J_1^4] \lesssim 2^4n^{-24}(k_2-k_1)^8n^{16}\E[R_{1}^{-4}]^2\E[R_{2}^{4}]^2  = 2^4\E[R_{1}^{-4}]^2\E[R_{2}^{4}]^2(k_2/n - k_1/n)^8.
		\end{align*}
		We repeatedly apply the H\"older's inequality and the above bound for the expectation, and we have $\E[J_s^4] \lesssim (k_2/n - k_1/n)^8$ for $s = 1,3,5,8$ since there are 8 summations in each $\E[J_s^4]$ which take the sum from $k_1 + 1$ to $k_2$, and $\E[J_s^4] \lesssim (k_2/n - k_1/n)^4$ for $s = 2,4,6,7,9,10$ since there are only 4 summations in each $\E[J_s^4]$ which take the sum from $k_1 + 1$ to $k_2$. Combining these results we have $I_1 \lesssim (k_2/n - k_1/n)^4.$        
		
		We can also show $I_2 \lesssim (l_2/n - l_1/n)^4$, and $I_3 \lesssim (m_2/n - m_1/n)^4$. Since the steps are very similar to the the arguments for $I_1$, we omit the details here. Thus, for any $u = (u_1,u_2,u_3), v = (v_1,v_2,v_3) \in [0,1]^3$, we have
		$$\E[(G_n^{(\mathcal{R}_n,s)}(u) - G_n^{(\mathcal{R}_n,s)}(v))^8] \leq C'((\floor{nu_1}/n - \floor{n v_1}/n)^4 + (\floor{nu_2}/n - \floor{n v_2}/n)^4 + (\floor{nu_3}/n - \floor{n v_3}/n)^4),$$
		for some positive constant $C' > 0$. It is easy to see that 
		\begin{align*}
			(\floor{nu_1}/n - \floor{n v_1}/n)^4 &= ((u_1 - v_1) - (\{u_1\} - \{v_1\})/n)^4 \lesssim (u_1 - v_1)^4 + (\{u_1\} - \{v_1\})^4/n^4\\
			&\lesssim (u_1 - v_1)^4 + 1/n^4.
		\end{align*}
		So 
		\begin{align*}
			\E[(G_n^{(\mathcal{R}_n,s)}(u) - G_n^{(\mathcal{R}_n,s)}(v))^8] &\leq C((u_1 - v_1)^4 + (u_2 - v_2)^4 + (u_3 - v_3)^4 + 1/n^4) \\
			&= C(\|u-v\|_4^4 + 1/n^4) \leq C(\|u-v\|^4 + 1/n^4),
		\end{align*}
		since $\|u-v\|_4^4 = \sum_{i = 1}^3(u_i - v_i)^4 \leq \sum_{i,j = 1}^3(u_i - v_i)^2(u_j - v_j)^2 = (\sum_{i}^3(u_i - v_i)^2)^2 = \|u-v\|^4.$ This completes the proof of tightness. 
	}
	\qed
	\vspace{4mm}

	\subsection{Proof of Theorem \ref{thm_rsrmpower}}
	
	(i)
	Under Assumption \ref{ass_power}, conditional on $\mathcal{R}_n$, we still have almost surely
	$$
	\frac{1}{p}\|Y_i-Y_j\|^2=\frac{1}{p}\sum_{k=1}^p \left(\frac{X_{i,k}}{R_i}-\frac{X_{j,k}}{R_j}\right)^2+\frac{2}{p}(\mu_i-\mu_j) \left(\frac{X_{i,k}}{R_i}-\frac{X_{j,k}}{R_j}\right)+\frac{1}{p}\|\mu_i-\mu_j\|^2\to \sigma^2(R_i^{-2}+R_j^{-2})
	$$
	as  conditional on $\mathcal{R}_n$,  $\{R_i^{-1}X_{i,k}\}_{k=1}^p$   is still a $\rho$-mixing sequence. 
	
	Recall  (\ref{D34}), conditional on $\mathcal{R}_n$, we mainly work on $D_3(k;l,m)$ since $D_4(k;l,m)$ is of a smaller order, where 
	$$
	D_3(k;l,m)\\=\sum_{\substack{l\leq j_1,j_3\leq k\\j_1\neq j_3}}\sum_{\substack{k+1\leq j_2,j_4\leq m\\j_2\neq j_4}}\frac{(Y_{j_1}-Y_{j_2})'(Y_{j_3}-Y_{j_4})}{(R_{j_1}^{-2}+R_{j_2}^{-2})^{1/2}(R_{j_3}^{-2}+R_{j_4}^{-2})^{1/2}}.$$
	
	By symmetry, we only consider the case $l<k\leq k^*<m$, and the summation in $D_3(k;l,m)$ can be decomposed into
	$$
	\sum_{\substack{l\leq j_1,j_3\leq k\\j_1\neq j_3}}\sum_{\substack{k+1\leq j_2,j_4\leq m\\j_2\neq j_4}}=\sum_{\substack{l\leq j_1,j_3\leq k\\j_1\neq j_3}}\Big\{\sum_{\substack{k+1\leq j_2,j_4\leq k^*\\j_2\neq j_4}} +\sum_{\substack{k^*+1\leq j_2,j_4\leq m\\j_2\neq j_4}}+\sum_{j_2=k+1}^{k^*}\sum_{j_4=k^*+1}^{m}+\sum_{j_4=k+1}^{k^*}\sum_{j_2=k^*+1}^{m}\Big\},
	$$
	according to the relative location of $j_2, j_4$ and $k^*$. 
	
	Then, it is not hard to see that 
	\begin{flalign*}
		D_3(k;l,m)=&\sum_{\substack{l\leq j_1,j_3\leq k\\j_1\neq j_3}}\sum_{\substack{k+1\leq j_2,j_4\leq m\\j_2\neq j_4}}\frac{(X_{j_1}/R_{j_1}-X_{j_2}/R_{j_2})'(X_{j_3}/R_{j_3}-X_{j_4}/R_{j_4})}{(R_{j_1}^{-2}+R_{j_2}^{-2})^{1/2}(R_{j_3}^{-2}+R_{j_4}^{-2})^{1/2}}
		\\&+\sum_{\substack{l\leq j_1,j_3\leq k\\j_1\neq j_3}}\sum_{\substack{k^*+1\leq j_2,j_4\leq m\\j_2\neq j_4}}\frac{\|\delta\|^2}{(R_{j_1}^{-2}+R_{j_2}^{-2})^{1/2}(R_{j_3}^{-2}+R_{j_4}^{-2})^{1/2}}\\&-
		\sum_{\substack{l\leq j_1,j_3\leq k\\j_1\neq j_3}}\sum_{j_2=k^*+1}^{m}\sum_{j_4=k+1, j_4\neq j_2}^{m} \frac{\delta'(X_{j_1}/R_{j_1}-X_{j_2}/R_{j_2})}{(R_{j_1}^{-2}+R_{j_2}^{-2})^{1/2}(R_{j_3}^{-2}+R_{j_4}^{-2})^{1/2}}\\&-\sum_{\substack{l\leq j_1,j_3\leq k\\j_1\neq j_3}}\sum_{j_4=k^*+1}^{m}\sum_{j_2=k+1, j_4\neq j_2}^{m} \frac{\delta'(X_{j_3}/R_{j_3}-X_{j_4}/R_{j_4})}{(R_{j_1}^{-2}+R_{j_2}^{-2})^{1/2}(R_{j_3}^{-2}+R_{j_4}^{-2})^{1/2}}\\
		:=&\sum_{i=1}^4D_{3,i}(k;l,m).
	\end{flalign*}
	Under Assumption \ref{ass_power}, and conditional on $\mathcal{R}_n$, similar to (\ref{D3}), we can show  for $i=3,4$.
	$$
	n^{-3}\|\Sigma\|_F^{-1}D_{3,i}(k;l,m)=o_p(1),
	$$
	while by (\ref{D3_con}), $n^{-3}\|\Sigma\|_F^{-1}D_{3,1}(k;l,m)\overset{\mathcal{D}}{\rightarrow} G^{(\mathcal{R}_n,s)}(k/n;l/n,m/n)$.

	Hence,  if $n\E(R^{-2})^{-1}\|\Sigma\|_F^{-1}\|\delta\|^2\to c_n$ as $p\to\infty$, then conditional on $\mathcal{R}_n$,  we obtain that
	\begin{flalign*}
		&n^{-3}\|\Sigma\|_F^{-1}p\sigma^2D^{(s)}(k;l,m)=n^{-3}\|\Sigma\|_F^{-1}D_3(k;l,m)+o_p(1)\\=&n^{-3}\|\Sigma\|_F^{-1}[D_{3,1}(k;l,m)+D_{3,2}(k;l,m)]+o_p(1)\\=&n^{-3}\|\Sigma\|_F^{-1}\sum_{\substack{l\leq j_1,j_3\leq k\\j_1\neq j_3}}\sum_{\substack{k+1\leq j_2,j_4\leq m\\j_2\neq j_4}}\frac{(X_{j_1}/R_{j_1}-X_{j_2}/R_{j_2})'(X_{j_3}/R_{j_3}-X_{j_4}/R_{j_4})}{(R_{j_1}^{-2}+R_{j_2}^{-2})^{1/2}(R_{j_3}^{-2}+R_{j_4}^{-2})^{1/2}}
		\\&+n^{-4}\sum_{\substack{l\leq j_1,j_3\leq k\\j_1\neq j_3}}\sum_{\substack{k^*+1\leq j_2,j_4\leq m\\j_2\neq j_4}}\frac{n\|\Sigma\|_F^{-1}\|\delta\|^2}{(R_{j_1}^{-2}+R_{j_2}^{-2})^{1/2}(R_{j_3}^{-2}+R_{j_4}^{-2})^{1/2}}+o_p(1)\\
		\overset{\mathcal{D}}{\rightarrow}&G^{(\mathcal{R}_n,s)}(k/n;l/n,m/n)+c_n\Delta_n^{(\mathcal{R}_n,s)}(k/n;l/n,m/n)
	\end{flalign*}
	where 
	\begin{flalign}\label{DRSRM}
		\begin{split}
			&\Delta_n^{(\mathcal{R}_n,s)}(k/n;l/n,m/n)\\=&\begin{cases}
				n^{-4}\sum_{\substack{l\leq j_1,j_3\leq k\\j_1\neq j_3}}\sum_{\substack{k^*+1\leq j_2,j_4\leq m\\j_2\neq j_4}}\frac{\E(R^{-2})}{(R_{j_1}^{-2}+R_{j_2}^{-2})^{1/2}(R_{j_3}^{-2}+R_{j_4}^{-2})^{1/2}}, \quad & l<k\leq k^*<m \\
				n^{-4}\sum_{\substack{l\leq j_1,j_3\leq k^*\\j_1\neq j_3}}\sum_{\substack{k+1\leq j_2,j_4\leq m\\j_2\neq j_4}}\frac{\E(R^{-2})}{(R_{j_1}^{-2}+R_{j_2}^{-2})^{1/2}(R_{j_3}^{-2}+R_{j_4}^{-2})^{1/2}},  \quad & l<k^*<k<m\\
				0, \quad & otherwise.
			\end{cases}
		\end{split}
	\end{flalign}

	Hence, we have 
	\begin{flalign*}
		&T_n^{(s)}|\mathcal{R}_n\overset{\mathcal{D}}{\rightarrow}\mathcal{T}_n^{(\mathcal{R}_n,s)}(c_n,\Delta_n^{(\mathcal{R}_n,s)}):=\sup_{k=4,\cdots,n-4}\\&
		\frac{n[G_n^{(\mathcal{R}_n,s)}(\frac{k}{n};\frac{1}{n},1)+c_n\Delta_n^{(\mathcal{R}_n,s)}(\frac{k}{n};\frac{1}{n},1)]^2}{\sum_{t=2}^{k-1}[G_n^{(\mathcal{R}_n,s)}(\frac{t}{n};\frac{1}{n},\frac{k}{n})+c_n\Delta_n^{(\mathcal{R}_n,s)}(\frac{t}{n};\frac{1}{n},\frac{k}{n})]^2+\sum_{t=k+2}^{n-2}[G_n^{(\mathcal{R}_n,s)}(\frac{t}{n};\frac{(k+1)}{n},1)+c_n\Delta_n^{(\mathcal{R}_n,s)}(\frac{t}{n};\frac{(k+1)}{n},1)]^2}.
	\end{flalign*}
	
	For $T_n$, by similar arguments as above, we have 
	\begin{flalign*}
		n^{-3}\|\Sigma\|_F^{-1}D(k;l,m)=&n^{-3}\|\Sigma\|_F^{-1}\sum_{\substack{l\leq j_1,j_3\leq k\\j_1\neq j_3}}\sum_{\substack{k+1\leq j_2,j_4\leq m\\j_2\neq j_4}}(X_{j_1}/R_{j_1}-X_{j_2}/R_{j_2})'(X_{j_3}/R_{j_3}-X_{j_4}/R_{j_4})
		\\&+n^{-4}\sum_{\substack{l\leq j_1,j_3\leq k\\j_1\neq j_3}}\sum_{\substack{k^*+1\leq j_2,j_4\leq m\\j_2\neq j_4}}n\|\Sigma\|_F^{-1}\|\delta\|^2+o_p(1)\\
		\overset{\mathcal{D}}{\rightarrow}&G^{(\mathcal{R}_n)}(k/n;l/n,m/n)+c_n\E(R^{-2})\Delta_n(k/n;l/n,m/n).
	\end{flalign*}
	Hence, we have 
	\begin{flalign*}
		&T_n^{(s)}|\mathcal{R}_n\overset{\mathcal{D}}{\rightarrow}\mathcal{T}_n^{(\mathcal{R}_n)}(c_n,\Delta_n^{(\mathcal{R}_n,s)}):=\sup_{k=4,\cdots,n-4}\\&
		\frac{n[G_n^{(\mathcal{R}_n)}(\frac{k}{n};\frac{1}{n},1)+c_n\E(R^{-2})\Delta_n(\frac{k}{n};\frac{1}{n},1)]^2}{\sum_{t=2}^{k-1}[G_n^{(\mathcal{R}_n)}(\frac{t}{n};\frac{1}{n},\frac{k}{n})+c_n\E(R^{-2})\Delta_n(\frac{t}{n};\frac{1}{n},\frac{k}{n})]^2+\sum_{t=k+2}^{n-2}[G_n^{(\mathcal{R}_n)}(\frac{t}{n};\frac{(k+1)}{n},1)+c_n\E(R^{-2})\Delta_n(\frac{t}{n};\frac{(k+1)}{n},1)]^2}.
	\end{flalign*}

	(ii)
	Note that for any $u=(u_1,u_2,u_3)^{\top}\in[0,1]^3$ such that $u_2\leq u_1\leq u_3$, as $n\to\infty$,
	$$
	\Delta_n^{(\mathcal{R}_n,s)}(\lfloor nu_1\rfloor/n;\lfloor nu_2\rfloor/n,\lfloor nu_3\rfloor/n)\to_p\E(R_1^{-2})E^2\Big[\frac{R_1R_2}{\sqrt{R_1^{2}+R_2^2}}\Big]\Delta(u_1;u_2,u_3).
	$$
	by the law of large numbers for $U$-statistics (since $\Delta_n^{(\mathcal{R}_n,s)}$ can be viewed as a two sample $U$-statistic).  Then using the similar arguments  in the proof of Theorem \ref{thm_rsrm}  (ii), we have 
	$$
	\Delta_n^{(\mathcal{R}_n,s)}(\cdot)\rightsquigarrow \E(R_1^{-2})E^2\Big[\frac{R_1R_2}{\sqrt{R_1^{2}+R_2^2}}\Big]\Delta(\cdot).
	$$

	Note that $\Delta(\cdot)$ is deterministic, and  recall $G_n^{(\mathcal{R}_n,s)}(\cdot)\rightsquigarrow \E \Big[\frac{R_1R_2}{\sqrt{(R_1^2+R_3^2)(R_2^2+R_3^2)}}\Big]\sqrt{2}G(\cdot)$ in the proof of Theorem \ref{thm_rsrm} (ii), by similar arguments in the proof of Theorem 3.6 in \cite{wang2021inference}, we have 
	$$
	G_n^{(\mathcal{R}_n,s)}(\cdot)+c_n\Delta_n^{(\mathcal{R}_n,s)}(\cdot)\rightsquigarrow \E\Big[\frac{R_1R_2}{\sqrt{(R_1^2+R_3^2)(R_2^2+R_3^2)}}\Big]\sqrt{2}G(\cdot)+c\E(R_1^{-2})E^2\Big[\frac{R_1R_2}{\sqrt{R_1^{2}+R_2^2}}\Big]\Delta(\cdot).
	$$
	Similarly, 
	$$
	G_n^{(\mathcal{R}_n)}(\cdot)+c_n\E(R^{-2})\Delta_n^{(\mathcal{R}_n,s)}(\cdot)\rightsquigarrow \E(R^{-2})\sqrt{2}G(\cdot)+c\E(R_1^{-2})\Delta(\cdot).
	$$
	The result follows by the continuous mapping theorem. Here, the multiplicative $K>1$ follows by the proof of Theorem 3.2 in \cite{chakraborty2017tests}.

	\qed
	\vspace{4mm}

	\section{Auxiliary Lemmas}\label{sec:lemma}
	\begin{lemma}\label{lem_fix}
		Under Assumptions \ref{ass_model} and \ref{ass_mixing},  let $n\geq 8$ be a fixed number, and for any $0\leq k<m\leq n$, let $Z(k,m)=\sum_{i=k+1}^{m}\sum_{j=k}^{i-1}X_i'X_j.$ Then,  as $p\to \infty$,
		$$\frac{\sqrt{2}}{n\|\Sigma\|_F}Z(k,m)\overset{\mathcal{D}}{\rightarrow} Q_n(\frac{k}{n},\frac{m}{n}),$$
		where 
		$Q_n(a,b)$ is a centered Gaussian process defined on $[0,1]^2$ with covariance structure given by:
		\begin{flalign*}
			&\mathrm{Cov}(Q_n(a_1,b_1),Q_n(a_2,b_2))\\=&n^{-2}( \lfloor nb_1\rfloor\land \lfloor nb_2\rfloor-\lfloor na_1\rfloor\lor \lfloor na_2\rfloor)(\lfloor nb_1\rfloor\land \lfloor nb_2\rfloor-\lfloor na_1\rfloor\lor\lfloor na_2\rfloor+1)\mathbf{1}(b_1\land b_2>a_1\lor a_2).
		\end{flalign*}
		
	\end{lemma}
	\textsc{Proof of Lemma \ref{lem_fix}}
	
	By Cram\'er-Wold device, it suffices to show that for fixed $n$ and $N$, any sequences of $\{\alpha_i\}_{i=1}^N$, $\alpha_i\in\mathbb{R}$,
	$$
	\sum_{i=1}^{N}\alpha_i \frac{\sqrt{2}}{n\|\Sigma\|_F}Z(k_i,m_i)\overset{\mathcal{D}}{\rightarrow} \sum_{i=1}^{N}\alpha_iQ_n(\frac{k_i}{n},\frac{m_i}{n}),
	$$
	where $1\leq k_i\leq m_i\leq n$ are integers.
	
	For simplicity, we consider the case of $N=2$, and by symmetry there are basically three types of enumerations of $(k_1,m_1,k_2,m_2)$:  (1) $k_1\leq m_1\leq k_2\leq m_2$; (2) $k_1\leq k_2\leq m_1\leq m_2$; (3) $k_1\leq k_2\leq m_2\leq m_1$.  
	
	Define $\xi^{(1)}_{i,t}=X_{i,t}\sum_{j=k_1}^{i-1}X_{j,t}$, and $\xi^{(2)}_{i,t}=X_{i,t}\sum_{j=k_2}^{i-1}X_{j,t}$. Then, we can show 
	\begin{flalign*}
		&\frac{\sqrt{2}}{n\|\Sigma\|_F}[\alpha_1Z(k_1,m_1)+\alpha_2Z(k_2,m_2)]
		\\=&\frac{\sqrt{2}}{n\|\Sigma\|_F}\Big(\alpha_1\sum_{i=k_1+1}^{m_1}\sum_{j=k_1}^{i-1}X_i'X_j+\alpha_2\sum_{i=k_2+1}^{m_2}\sum_{j=k_2}^{i-1}X_i'X_j\Big)
		\\=&\left\{\begin{aligned}
			&\frac{\sqrt{2}}{n\|\Sigma\|_F}\sum_{t=1}^{p}\Big(\sum_{i=k_1+1}^{m_1}\alpha_1\xi^{(1)}_{i,t}+\sum_{i=k_2+1}^{m_2}\alpha_2\xi^{(2)}_{i,t}\Big),\quad &\text{Case (1)}\\
			&\frac{\sqrt{2}}{n\|\Sigma\|_F}\sum_{t=1}^{p}\Big(\sum_{i=k_1+1}^{k_2}\alpha_1\xi^{(1)}_{i,t}+\sum_{i=k_2+1}^{m_1}[\alpha_1\xi^{(1)}_{i,t}+\alpha_2\xi^{(2)}_{i,t}]+\sum_{i=m_1+1}^{m_2}\alpha_2\xi^{(2)}_{i,t}\Big),\quad &\text{Case (2)}\\
			&\frac{\sqrt{2}}{n\|\Sigma\|_F}\sum_{t=1}^{p}\Big(\sum_{i=k_1+1}^{k_2}\alpha_1\xi^{(1)}_{i,t}+\sum_{i=k_2+1}^{m_2}[\alpha_1\xi^{(1)}_{i,t}+\alpha_2\xi^{(2)}_{i,t}]+\sum_{i=m_2+1}^{m_1}\alpha_1\xi^{(1)}_{i,t}\Big),\quad &\text{Case (3)}\\
		\end{aligned}\right.
	\end{flalign*}
	For simplicity, we consider the Case (2), and using the independence of $X_i$, one can show that $S_1=\frac{\sqrt{2}}{n\|\Sigma\|_F}\sum_{t=1}^{p}\sum_{i=k_1+1}^{k_2}\alpha_1\xi^{(1)}_{i,t}$, $S_2=\frac{\sqrt{2}}{n\|\Sigma\|_F}\sum_{t=1}^{p}[\alpha_1\xi^{(1)}_{i,t}+\alpha_2\xi^{(2)}_{i,t}]$ and $S_3=\frac{\sqrt{2}}{n\|\Sigma\|_F}\sum_{t=1}^{p}\sum_{i=k_2+1}^{m_2}\alpha_2\xi^{(2)}_{i,t}$ are independent. Then by Theorem 4.0.1 in Lin and Lu (2010), they are asymptotically normal with variances  given by 
	\begin{flalign*}
		\mathrm{Var}(S_1)=&n^{-2}\alpha_1^2(k_2-k_1)(k_2-k_1+1),\\
		\mathrm{Var}(S_2)=&n^{-2}[\alpha_1^2(m_1-k_2)(k_2-k_1+1+m_1-k_1)+2\alpha_1\alpha_2(m_1-k_2)(m_1-k_2+1)\\&+\alpha_2^2(m_1-k_2)(m_1-k_2+1)],\\
		\mathrm{Var}(S_1)=&n^{-2}\alpha_2^2(m_2-m_1)(m_2-k_2+m_1-k_2+1).
	\end{flalign*}
	Similarly, we can obtain the asymptotic normality for Case (1) and Case (3).
	
	Hence, 
	$$
	\frac{\sqrt{2}}{n\|\Sigma\|_F}[\alpha_1Z(k_1,m_1)+\alpha_2Z(k_2,m_2)]\overset{\mathcal{D}}{\rightarrow} N(0,\frac{\tau^2}{n^2 }),
	$$
	where $$
	\tau^2=\begin{cases}
		\alpha_1^2{(m_1-k_1)(m_1-k_1+1)}+\alpha_2^2{(m_2-k_2)(m_2-k_2+1)},\quad& \text{Case (1)}\\
		\alpha_1^2{(m_1-k_1)(m_1-k_1+1)}+\alpha_2^2{(m_2-k_2)(m_2-k_2+1)}+2\alpha_1\alpha_2(m_1-k_2)(m_1-k_2+1),\quad &\text{Case (2)}\\
		\alpha_1^2{(m_1-k_1)(m_1-k_1+1)}+\alpha_2^2{(m_2-k_2)(m_2-k_2+1)}+2\alpha_1\alpha_2(m_2-k_2)(m_2-k_2+1),\quad &\text{Case (3)}.
	\end{cases}
	$$
	Hence, the case of $N=2$ is proved by examining the covariance structure of $Q_n$ defined in Theorem \ref{thm_main}. The cases $N>2$ are similar. 
	\qed
	\vspace{4mm}

	\begin{lemma}\label{lem_equiv}
		As $n\to\infty$, we have for any $0\leq a_1<r_1<b_1\leq 1$ and $0\leq a_2<r_2<b_2\leq 1$, as $n \rightarrow \infty$,
		\begin{flalign*}
			&\mathrm{Cov}(G_n^{(\mathcal{R}_n,s)}(r_1;a_1,b_1),G_n^{(\mathcal{R}_n,s)}(r_2;a_2,b_2))\\
			\rightarrow_p&2\E^{2}\Big[\frac{R_1R_2}{\sqrt{(R_1^2+R_3^2)(R_2^2+R_3^2)}}\Big]\mathrm{Cov}(G(r_1;a_1,b_1),G(r_2;a_2,b_2)).
		\end{flalign*}
		
	\end{lemma}
	
	There are 9 terms in the  covariance structure given in (\ref{cov}), for first one,  we have
	\begin{flalign*}
		&2n^{-6}\sum_{\substack{\lfloor n(a_1\lor a_2)\rfloor\leq j_1,j_2\leq \lfloor n(r_1\land r_2)\rfloor\\j_1\neq j_2}}R_{j_1}^{-2}R_{j_2}^{-2}A_{j_1,j_2}(\lfloor nr_1\rfloor;\lfloor na_1\rfloor,\lfloor nb_1\rfloor)A_{j_1,j_2}(\lfloor nr_2\rfloor;\lfloor na_2\rfloor,\lfloor nb_2\rfloor)
		\\=2&n^{-6}\sum_{\substack{\lfloor n(a_1\lor a_2)\rfloor\leq j_1,j_2\leq \lfloor n(r_1\land r_2)\rfloor\\j_1\neq j_2}}R_{j_1}^{-2}R_{j_2}^{-2}\sum_{\substack{\lfloor nr_1\rfloor+1\leq j_3,j_4\leq \lfloor nb_1\rfloor\\j_3\neq j_4}}\frac{R_{j_1}R_{j_3}}{\sqrt{(R_{j_1}^2+R_{j_3}^2)}}\frac{R_{j_2}R_{j_4}}{\sqrt{(R_{j_2}^2+R_{j_4}^2)}}\\&\times\sum_{\substack{\lfloor nr_2\rfloor+1\leq j_5,j_6\leq \lfloor nb_2\rfloor\\j_5\neq j_6}}\frac{R_{j_1}R_{j_5}}{\sqrt{(R_{j_1}^2+R_{j_5}^2)}}\frac{R_{j_2}R_{j_6}}{\sqrt{(R_{j_2}^2+R_{j_6}^2)}}
		\\=2&n^{-2}\sum_{\substack{\lfloor n(a_1\lor a_2)\rfloor\leq j_1,j_2\leq \lfloor n(r_1\land r_2)\rfloor\\j_1\neq j_2}}R_{j_1}^{-2}R_{j_2}^{-2}(b_1-r_1)^2\Big\{\E\Big[\frac{R_{j_1}R_{j_3}}{\sqrt{(R_{j_1}^2+R_{j_3}^2)}}\frac{R_{j_2}R_{j_4}}{\sqrt{(R_{j_2}^2+R_{j_4}^2)}}|R_{j_1},R_{j_2}\Big]+o_p(1)\Big\}\\&\times(b_2-r_2)^2\Big\{\E\Big[\frac{R_{j_1}R_{j_5}}{\sqrt{(R_{j_1}^2+R_{j_6}^2)}}\frac{R_{j_2}R_{j_6}}{\sqrt{(R_{j_2}^2+R_{j_6}^2)}}|R_{j_1},R_{j_2}\Big]+o_p(1)\Big\}
		\\\to_p& 2[(r_1\land r_2)-(a_1\lor a_2)]^2(b_1-r_1)^2(b_2-r_2)^2 \E^2\Big[\frac{R_1R_2}{\sqrt{(R_1^2+R_3^2)(R_2^2+R_3^2)}}\Big].
	\end{flalign*} 
	where the last equality holds by applying the law of large numbers for $U$-statistics to $R_{j_3}, R_{j_4}$ and $R_{j_5},R_{j_6}$, and the last holds by the law of large numbers of $U$-statistics to $R_{j_1}, R_{j_2}$.
	
	Therefore, similar arguments for other terms indicate that 
	\begin{flalign*}
		&2^{-1}\E^{-2}\Big[\frac{R_1R_2}{\sqrt{(R_1^2+R_3^2)(R_2^2+R_3^2)}}\Big]\lim\limits_{n\to\infty}\mathrm{Cov}(G_n^{(\mathcal{R}_n,s)}(\lfloor nr_1\rfloor;\lfloor na_1\rfloor,\lfloor nb_1\rfloor),G_n^{(\mathcal{R}_n,s)}(\lfloor nr_2\rfloor;\lfloor na_2\rfloor,\lfloor nb_2\rfloor))
		\\=&[(r_1\land r_2)-(a_1\lor a_2)]^2(b_1-r_1)^2(b_2-r_2)^2\mathbf{1}((r_1\land r_2)>(a_1\lor a_2))\\&+[(r_1\land b_2)-(a_1\lor r_2)]^2(b_1-r_1)^2(r_2-a_2)^2\mathbf{1}((r_1\land b_2)>(a_1\lor r_2))\\&-4[r_2-(a_1\lor a_2)][(b_2\land r_1)-r_2](b_1-r_1)^2(b_2-r_2)(r_2-a_2)\mathbf{1}(r_1>r_2,r_2>(a_1\lor a_2),(b_2\land r_1)>r_2)
		\\&+[(b_1\land r_2)-(r_1\lor a_2)]^2(r_1-a_1)^2(b_2-r_2)^2\mathbf{1}((b_1\land r_2)>(r_1\lor a_2))\\&+[(b_1\land b_2)-(r_1\lor r_2)]^2(r_1-a_1)^2(r_2-a_2)^2\mathbf{1}((b_1\land b_2)-(r_1\lor r_2))\\
		&-4[r_2-(r_1\lor a_2)][(b_1\land b_2)-r_2](r_2-a_2)(b_2-r_2)(r_1-a_1)^2\mathbf{1}(b_1>r_2,r_2>(r_1\lor a_2),(b_1\land b_2)>r_2)
		\\&-4[r_1-(a_1\lor a_2)][(b_1\land r_2)-r_1](r_1-a_1)(b_1-r_1)(b_2-r_2)^2\mathbf{1}(r_2>r_1,r_1>(a_1\lor a_2),(b_1\lor r_2)>r_1)
		\\&-4[r_1-(r_2\lor a_1)][(b_1\land b_2)-r_1](r_1-a_1)(b_1-r_1)(r_2-a_2)^2\mathbf{1}(b_2>r_1,r_1>(r_2\lor a_1),(b_1\land b_2)>r_1)
		\\&+4[(r_1\land r_2)-(a_1\lor a_2)][(b_1\land b_2)-(r_1\land r_2)](r_1-a_1)(b_1-r_1)(r_2-a_2)(b_2-r_2)\\&\times\mathbf{1}((r_1\land r_2)>(a_1\lor a_2),(b_1\land b_2)>(r_1\land r_2)).
	\end{flalign*}
	
	This is indeed the covariance structure of $G(\cdot)$ after tedious algebra.
	\qed
	\vspace{4mm}
	
	\bibliographystyle{apalike}
	\bibliography{reference}

\end{document}